\documentclass[12pt, preprint]{aastex}
\usepackage{bm, graphicx, subfigure, amsmath, morefloats}

\newcommand{\project}[1]{\textsl{#1}}
 
\newcommand{\tc}{\project{The~Cannon}} 
\newcommand{\apogee}{\project{\textsc{apogee}}}
\newcommand{\apokasc}{\project{\textsc{apokasc}}}
\newcommand{\aspcap}{\project{\textsc{aspcap}}}
\newcommand{\corot}{\project{Corot}}
\newcommand{\kepler}{\project{Kepler}}

\newcommand{\gaiaeso}{\project{Gaia--\textsc{eso}}}
\newcommand{\galah}{\project{\textsc{galah}}}
\newcommand{\most}{\project{\textsc{most}}}
\newcommand{\code}[1]{\texttt{#1}}

\newcommand{\teff}{\mbox{$\rm T_{eff}$}}

\newcommand{\feh}{\mbox{$\rm [Fe/H]$}}
\newcommand{\xfe}{\mbox{$\rm [X/Fe]$}}
\newcommand{\alphafe}{\mbox{$\rm [\alpha/Fe]$}}

\newcommand{\logg}{\mbox{$\rm \log g$}}

\newcommand{\rgal}{\mbox{$\rm R_{GAL}$}}

\newcommand{\set}[1]{\bm{#1}}
\newcommand{\starlabel}{\ell}
\newcommand{\starlabelvec}{\set{\starlabel}}

\newcommand{\numax}{$\nu_{\max}$}
\newcommand{\deltanu}{$\Delta\nu$}
\begin{document}

\title{Spectroscopic determination of masses (and implied ages) \\for red giants}
\author{M.~Ness\altaffilmark{1},
        David~W.~Hogg\altaffilmark{1,2,3},
        H.-W.~Rix\altaffilmark{1},
        M.~Martig \altaffilmark{1}, 
     Marc~H.~Pinsonneault\altaffilmark{4},
        A.Y.~Q~Ho \altaffilmark{1, 5}}
\altaffiltext{1}{Max-Planck-Institut f\"ur Astronomie, K\"onigstuhl 17, D-69117 Heidelberg, Germany}
\altaffiltext{2}{Center for Cosmology and Particle Physics, Department of Phyics, New York University, 4 Washington Pl., room 424, New York, NY 10003, USA}
\altaffiltext{3}{Center for Data Science, New York University, 726 Broadway, 7th Floor, New York, NY 10003, USA}
\altaffiltext{4}{Department of Astronomy, The Ohio State University, Columbus, OH 43210, USA; Center for Cosmology and Astroparticle Physics, The Ohio State University, Columbus, OH 43210, USA}
\altaffiltext{5}{Department of Astronomy, California Institute of Technology, 1200 East California Blvd, Pasadena, CA 91125, USA}
\email{ness@mpia.de}

\begin{abstract}%
The mass of a star is arguably its most fundamental parameter.
For red giant stars, tracers luminous enough to be observed
across the Galaxy, mass implies a stellar evolution age.
It has proven to be extremely difficult to infer ages and masses directly from red giant spectra using existing methods.
From the \kepler\ and \apogee\ surveys, samples of several thousand
stars exist with high-quality spectra and asteroseismic masses.
Here we show that from these data we can build a data-driven spectral
model using \tc, which can determine stellar masses to $\sim 0.07$~dex
from \apogee\ \textsc{dr12} spectra of red giants; these imply age
estimates accurate to $\sim 0.2$~dex (40 percent).
We show that \tc\ constrains these ages foremost from spectral regions
with CN absorption lines, elements whose surface abundances reflect
mass-dependent dredge-up.
We deliver an unprecedented catalog of 80,000 giants (including 20,000
red-clump stars) with mass and age estimates, spanning the entire disk
(from the Galactic center to $R\sim 20$~kpc).
We show that the age information in the spectra is not simply a
corollary of the birth-material abundances \feh\ and \alphafe, and
that even within a mono-abundance population of stars, there are age
variations that vary sensibly with Galactic position.
Such stellar age constraints across the Milky Way open up new avenues
in Galactic archeology.
\end{abstract}


\section{Introduction}\label{sec:Intro}

Age-dating of stars is fundamental to understanding and reconstructing the formation and evolution history of the Milky Way. Independent measurements for both the elemental abundances and ages of an extensive set of stars across the Milky Way would be a powerful constraint on galaxy and also on chemical evolution (presuming the chemical information is derived from material from which the stars have formed).
Yet, as almost all stars are in equilibrium\footnote{Most stars not in equilibrium undergo periodic variations, such as pulsations, but without discernible secular evolution.}, age is not a quantity that can be directly measured. Instead, one must rely on measuring instantaneous stellar properties (or ``labels'') that correlate with age in a physically understood way, or one which can be calibrated \citep[see][for an excellent review]{soderblom2010}.
Inevitably, stellar age estimates involve some form of stellar evolution models, both for stars in clusters and for single field stars. 

For the most part, age estimates from spectroscopic surveys have been determined for stars before or just after their main-sequence turn-off. 
In that regime, stellar evolutionary isochrones are well separated (at a given metallicity), and for well-measured \teff, \logg\ and \feh, ages follow from isochrone matching. 
Such stellar parameters are typically derived from high-resolution spectroscopy, which delivers low associated errors on the parameters \citep[e.g.][]{Bensby2013, Casagrande2011, haywood2013, B2014}.  To date, the largest homogeneous data set of stellar ages in the galactic disk has been derived in this fashion, from the Geneva Cophenhagen Survey (GCS). Yet, all 16,682 main sequence stars from GCS are located in the immediate solar neighbourhood of $<0.1$~kpc \citep{nordstrom2004short}.
Recent analogous analyses \citep[e.g.][]{haywood2013, B2014} have pushed to greater distances, but still remain limited to essentially the solar radius.

To map stellar ages throughout the Milky Way, one needs more luminous stars, in evolutionary phases that are prevalent  across most ages and metallicities. Giant stars satisfy these criteria. They have the advantage that their luminosities and colors vary relatively little with age, which makes age biases in flux-limited samples weaker. Yet, this also means that giant stars isochrones of different ages nearly overlap, making it all but impossible to get precise ages from \teff , \logg\  and \feh\ measurements, unless we have tiny errors in these measurements and enormous confidence in the accuracy of stellar isochrones. For reference, consider a typical solar abundance red giant at \logg\ = 2 with an age of 5 Gyr.  For the PARSEC isochrones   \citep{Bressan2012}, age differences of +/- 2 Gyr correspond to changes in \teff\ at fixed \logg\ of only $\approx$ 10 K, compared to shifts of $\approx$ 50 K for a 0.10 dex difference in \feh.  Furthermore, core helium burning stars that have experienced significant prior mass loss, red clump stars, are located close in the HR Diagram to less evolved first ascent red giant branch stars.  Even if the observational data is exact, absolute comparisons to stellar isochrones are uncertain; the absolute \teff\ of theoretical models, for example, is highly sensitive to the assumed efficiency of convection, typically parameterized with a mixing length.  However, for basically all post main sequence stars, in particular stars on the red giant branch or in the red clump, the stellar mass should be a powerful constraint on the stars' age \citep[see e.g.][]{Martig2014}. In that case the challenge is reduced to estimating stellar masses for extensive samples of giant stars throughout the Galaxy; these masses then imply ages.

In recent years, asteroseismology surveys such as \most\ \citep{most2005}, \corot\ \citep{corot2009}, and \kepler\ \citep{Bedding2010}, have
been extremely successful in producing information
about stellar interiors and hence masses, in particular for giant stars. 
These missions operate by taking high-cadence, high-precision stellar
photometry over long, uninterrupted time intervals, in which stellar oscillation 
modes are visible in the Fourier domain. These modes are related to the density and mass of the stars.
At present, all these astroseismological surveys cover only a few directions in the sky, and hence a small portion of the Galaxy.

At the same time, there are a number of large spectroscopic surveys, such
as \apogee\ \citep{Majewski2012}, \gaiaeso\ \citep{Gilmore2012} and \galah\ \citep{Freeman2012}. These surveys are 
producing high signal-to-noise, high-resolution spectra
of hundreds of thousands of stars across the entire sky. These allow measurements of properties including \teff, \logg, \feh\ and \xfe\ for many elements. A star's surface abundances, in particular \feh\ and \alphafe\ hold clues to its age, because the plausibility of stars forming from material of a given abundance dramatically varies with time and radius throughout the galaxy: e.g. stars 
were far more likely to have formed from metal-poor but $\alpha$-enhanced ISM than they are now. But such age constraints arise from the properties of the birth material, not from the current properties of the star itself, and hence age estimates and chemical evolution of the interstellar medium are inevitably degenerate \citep[see e.g.,][]{Schonrich2009, Ch2002}.

The question then naturally arises how one can combine the information from these two types of surveys, information about the stellar interior and masses from seismology; and
stellar parameters and element-abundances from spectroscopy. For stars that have been observed by both kinds of surveys this can be done at the catalog level \citep{Martig2014}.

As stars evolve to the red giant branch they develop deep surface convection zones and dredge up nuclear processed material in their interiors in a mass dependent fashion \citep{Iben1967}.  Elements whose surface abundance is particularly sensitive to this phenomenon include the light elements Li, Be, B, C (in particular the ratio of C12/C13) and N; C and N are measured by surveys such as APOGEE.  There is also evidence, especially in metal-poor stars, for changes in CNO abundances along the giant branch \citep{Kraft1994}, which requires a mixing process not usually included in standard models whose origin is still uncertain \citep{A2012}.  However, \textit{empirical} correlations between stars of known mass and surface abundances can yield powerful insights even without detailed knowledge of the underlying physics.

Following this approach, Martig et al., (2015) have used red giants in the \apokasc\ sample of stars, for which seismic parameters are known from \kepler\ \citep{P2014} and \teff, \logg, \feh, \alphafe\  are \xfe\ are measured from \apogee\ spectra \citep{Ahn2014}. Martig et al., (2015) have found a tight correlation between the masses determined from the standard seismic scaling relations and the [C/N] measurement from \apogee.  They determine a model for stellar mass and age, as a function of C and N abundance measurements. 

In this paper, we set out to develop a data-driven and far-reaching connection between the asteroseismic and the spectroscopic results for giant stars, with the ultimate goal of determining stellar masses of giants, and hence ages, directly from spectra. The goal is to derive age estimates that do not simply reflect the abundances of the star's birth material (such as joint \feh\ and \alphafe\ estimates); we aim for age estimates that give meaningful results, {\sl even at a given [Fe/H] and [$\alpha$/Fe]}. We are however, with spectroscopic data, determining only the surface property of stars. The physical properties of mass and derived age can only be inferred, given theoretical expectations from stellar models between these physical quantities and stellar spectra.

For this purpose, we use a set of 1639 \apokasc\ stars from the \apogee\ DR12 spectral sample with stellar mass and \logg\ measurements from astroseismology \citep{P2014}, along with DR12's  \teff, \feh\ and \alphafe\ (see Figure \ref{fig:triangle} in the Appendix).  Using \tc~ \citep{Ness2015} we then generate a data-driven generative model for the five stellar labels \teff, \logg, \feh, \alphafe\ and mass. This model quantifies the information content at each pixel.  
Therefore, we can examine the origin of the information on these labels directly in the spectra. We have shown previously, in  \citet[][and Ho et al., in prep]{Ness2015}   that \tc\ is successful for delivering the labels of \teff, \logg, \feh\  and \alphafe\ for \apogee\ stars. With a training set of stars with know masses, we can expand the same approach to include a fifth stellar label, mass. With these data,  \tc\ is a direct framework to characterize the relationship between the surface spectroscopy and interior astroseismology, in order to jointly infer stellar properties, learning about stellar interiors from surface spectroscopy.

In Section \ref{sec:methods} we describe the implementation and application of \tc\ for the case at hand. We also lay out the verification of the mass label estimates, and we illustrate where in the spectrum the information on the various labels originates. 

We deliver our catalogue of mass and inferred ages for the $\approx$ 20,000 red clump stars in the \apogee\ survey, which have well known distances, as well as for the 60,000 red giant stars in the DR12 \apogee\ data release that are within the label (stellar-parameter) range of our training set.

\section{Methods and data}
\label{sec:methods}

\subsection{Implementation of \tc\ to include mass labels}

We make use of \tc\ \citep{Ness2015}, which is a data-driven method for
determining stellar parameters and abundances.
\tc\ is a probabilistic model of stellar spectra---meaning that it
produces a likelihood function or a probability density in spectral
space---that is itself a function of stellar parameters and chemical
abundances (which we collectively call ``labels'').
The model is not based on physical models, but is instead learned
from a training set of stars with (assumed) known labels.
This learning is called the ``training step''.
The model is used to label a new star \emph{not} in the training set,
by maximizing the likelihood of the label values given the new star's
spectrum.
This labeling is called the ``test step''.
\tc\ differs from standard machine-learning methods
(such as random forest or deep neural networks) in that it contains an
explicit likelihood function, at both the training step and the test
step, so it is able to account for heteroskedastic noise and missing
data in the spectra of both the training and test stars.

Generally, we take a spectral model to be characterized by a coefficient vector $\set{\theta}_\lambda$
that predicts the flux at every pixel $f_{n\lambda}$ for a given label vector $\starlabelvec_n$:

\begin{eqnarray}
f_{n\lambda} &=&
g(\starlabelvec_n |  \set{\theta}_\lambda) + \mbox{noise}
\label{eq:specmodel}\quad 
\end{eqnarray}

In detail, the likelihood function we use for \tc\ has a
Gaussian form at each measured spectral wavelength, with a mean that
is a quadratic function of the labels, and a variance that consists of
an intrinsic variance added to an observational noise variance (from
photon noise and other sources). For our model we use a the quadratic-in-labels form of \citet{Ness2015}. This model presumes that the continuum-normalized flux is a polynomial of the stellar labels, as below: 

$f_{n\lambda} =
\set{\theta}_\lambda^T \cdot \starlabelvec_n + \mbox{noise}$, 
but where $\set{\theta}_\lambda$ now contains 21 elements at every pixel. For the case of the five labels $(\teff , \logg , \feh, \alphafe, \mathcal{M})$ the label vector $\starlabelvec_n$
becomes  

\begin{equation}
\begin{split}
\starlabelvec_n \equiv  [1, \teff, \logg, \feh, \alphafe, \mathcal{M}, \teff^2, \\
 \teff\cdot\logg, \teff\cdot\feh, \teff\cdot\alphafe, \teff\cdot \mathcal{M},  \logg^2, \\
  \logg\cdot\feh, \logg\cdot\alphafe, \logg\cdot \mathcal{M}, \feh^2, \\
  \feh\cdot\alphafe, \feh\cdot \mathcal{M}, \alphafe^2, \alphafe\cdot\mathcal{M}, \mathcal{M}^2]
\end{split}
 \label{eq:quad}
\end{equation}

\subsection{Data}

We have shown in previous work \citep{Ness2015} that \tc\ does a good job
of modeling stellar spectra and delivering stellar parameters and
chemical abundances for stars with spectra taken by the \apogee\ project \citep{Majewski2012}.  \apogee\ is an SDSS \footnote{\url{www.sdss.org}}  \citep{E2011} infrared survey of the Milky Way disk, bulge and halo and has provided H-band spectra (1500-1700 nm) of about 150,000 stars in the public data release DR12.  
The three labels of \teff, \logg\ and \feh\ delivered with \tc\ were demonstrated in \citet{Ness2015}.  In this work we train on and then determine two additional labels, of \alphafe\ and mass. We train on log mass and infer the subsequent age, using stellar evolution models as described in Section 2.3. Our five labels are provided to \tc\ in the training step, and delivered by \tc\ in the test step.

\subsubsection{Training Data}

 The training set is comprised of 1639 stars taken from the \kepler\ field, the so called \apokasc\ sample \citep{P2014}  of stars observed by \apogee.
This sample of stars have high-quality infrared
spectra from \apogee\ and also asterosesmological measurements from the \kepler\ mission.
The \kepler\ Mission \citep{B2010} took continuous, 30-min cadence (or
higher cadence) photometric observations of more than $10^5$ stars,
providing (at least for giant stars) measurements of the
asteroseismological frequencies and frequency splittings that indicate
stellar interior density structure.
The two global asteroseismic parameters are the \numax\ and \deltanu\ quantities. These are the measurements from \kepler\ that indicate the interior structure of the star \citep[see][and references therein]{P2014}.
The asteroseismic measurements are used---with stellar models---to infer stellar masses and thus provide labels.

Our training set of 1639 \apokasc\ stars are described in \citet{Martig2014} and selected from the full \apokasc\ sample in \citet{P2014} based on additional quality cuts. This sample includes only stars with no warning or error in the \aspcap\ \code{FLAG} parameter provided by \apogee\ \citep{Ahn2014}, with no rotation flag set and with errors on the \deltanu\ and \numax\ less than 10\,percent are included. The \apokasc\ stars comprise a high signal-to-noise (SNR) sample, with an SNR $>$ 80. 

We work with the continuum-normalized DR12 spectra, and the method of continuum
estimation turns out to be important for performance. We use the aspcapStar files provided by \apogee, but apply our own signal-to-noise invariant continuum normalization by fitting a low-order polynomial to `true' continuum pixels, as described in \citet{Ness2015}. 
Five labels are used for training of \tc, the \teff, \logg, \feh, \alphafe\ and log mass. The label range of the training data set is shown in the Appendix, in Figure \ref{fig:triangle}. 
 The five training labels adopted are from the \aspcap-corrected values  \citep{Meszaros2013} for the \teff, \feh\ and \alphafe\ and the asteroseismic value for \logg, as determined from the measured $\nu_{max}$. The mass label was determined from the  \deltanu\ and \numax\ measurements using the standard seismic scaling relation \citep[e.g.][]{K1995}, as in Equation \ref{eq:mass}. 

\begin{equation} \label{eq:mass}
M= \left( \frac{\nu_{\mathrm{max}}}{\nu_{\mathrm{max,\odot}}}\right)^3\  \left( \frac{\Delta \nu}{\Delta \nu_{\odot}}\right)^{-4} \ \left( \frac{T_{\mathrm{eff}}}{T_{\mathrm{eff,\odot}}}\right)^{1.5} \ 
\end{equation}

We adopt  $T_{\mathrm{eff,\odot}}=5777$ K, $\nu_{\mathrm{max,\odot}}=3140\ \mu$Hz, $\Delta \nu_{\odot}=135.03\ \mu$Hz, as per \citet{Martig2014}. The solar values  $\Delta \nu_{\odot}$ and $\nu_{\mathrm{max,\odot}}$ are those used for the \apokasc\ catalogue and were obtained by \cite{Hekker2013}.

Note that modified scaling relations can be adopted in order to determine mass from the the asteroseismic parameters. \tc\ is a generalized method and in all cases, the results at the test step will be directly tied to the assumptions in the training step. \tc\ is implemented here as described in \citet{Ness2015} but using the model in Equation \ref{eq:quad}, with the mass label coming from the equation described in Equation \ref{eq:mass}.  

The scaling relationships rely on a combination of theoretically motivated and empirical arguments.  As such, their absolute values need to be calibrated by comparison with fundamental masses.  Radii are in reasonable agreement with parallax \citep{SA2012} and interferometry \citep{Huber2012} measurements.  However, there appear to be modest but real offsets between the expected and asteroseismic masses of open cluster red giants \citep{Brogaard2012} and somewhat larger ones for halo giants \citep{Epstein2014}.  These differences may depend on evolutionary state \citep{Miglio2012}, but are otherwise systematic rather than random in nature.  We therefore proceed with the masses as indicated by the unmodified relations, with the caution that there could be zero-point differences, metallicity-dependent stretches in the mass scale, and evolutionary-state dependent changes.  Despite these important caveats, we demonstrate that the relative masses inferred from these scalings produce sensible inferences about galactic properties.  It is straightforward to adopt corrected masses as new calibrations of the scaling relations arise.

\subsubsection{Test Data} 

We train out model using the \apokasc\ sample and then determine our five stellar labels of \teff, \logg, \feh, \alphafe\ and log mass for \apogee's DR12 red clump catalogue \citet{Bovy2014} and all red giants in \apogee's DR12 data release that are within the label range of our training dataset.  The test data is treated in the exact same way as the training data, as described in \citet{Ness2015} where we work with continuum normalized \apogee\ aspcapStar files and apply our own additional continuum normalization procedure. 

\subsection{From Masses to Ages}

Our asteroseismic calibration set measures only \textit{current} masses; a model is required to map these masses to their initial values.  In addition, the mapping of stellar mass to age depends on the adopted input physics Ð for example, the treatment of convective core overshooting for massive stars Ð as well as the detailed mixture of heavy elements and the assumed initial helium abundance \citep[see][for a detailed discussion]{soderblom2010}.  For red giants, the importance of mass loss depends sensitively on the luminosity and whether or not the star is a first ascent red giant or a red clump star.  Using the general formulation of \citet{R1975}, one would expect on dimensional grounds to have mass loss occur primarily on the upper red giant branch, when the surface gravity is low; there could also be mass loss associated with the ignition of helium in a degenerate medium.  Mass loss is therefore only likely to be important for red clump stars and for very luminous first ascent giants; the latter are rare in our sample.   

Globular cluster data requires modest (of order 0.2 solar masses) integrated mass loss on the giant branch with a stochastic dispersion of order 0.03 solar masses \citep[e.g.][]{Lee1990}.  The mass loss for higher metallicity red giants is less well-established, with some suggestion from Kepler data for a relatively weak mass loss \citep{Miglio2012}; note, however, that recent summaries of globular cluster data imply a larger scaling constant of order 0.48 \citep{McD2015}.  We therefore adopt a modest mass loss prescription ($\eta_{\mbox{Reimers}}$=0.2) to map current onto initial mass, with the caution that this may underestimate the effect for red clump stars.  For a recent discussion of the age uncertainties for red giants with asteroseismic masses see \citet{Casagrande2015}. 

 For our purposes, we are interested primarily in differential ages and in checking whether or not the usage of asteroseismic masses results in plausible age properties, not in rigorous absolute age measurements.  In the sections that follow, we explicitly distinguish between the ages of red giant and red clump stars to separate out the red clump sample whose ages depend on the assumptions concerning mass loss from the red giant sample that is relatively insensitive. (Also note that assuming a red clump evolutionary state for example, instead of a red giant evolutionary state for interpolating mass to age, does not change the age distribution of the sample. Individual stellar age differences are on the order of 5\,percent between these two evolutionary states).  Finally, we note that a real astrophysical sample will include both mergers of low mass stars and stars that have had their envelopes stripped by a companion; care must be taken in population modeling to distinguish such astrophysical backgrounds from very young or old populations respectively.

\section{Results}

\subsection{Validation of the mass determination}

\begin{figure}[h]
\centering
        \includegraphics[scale=0.4]{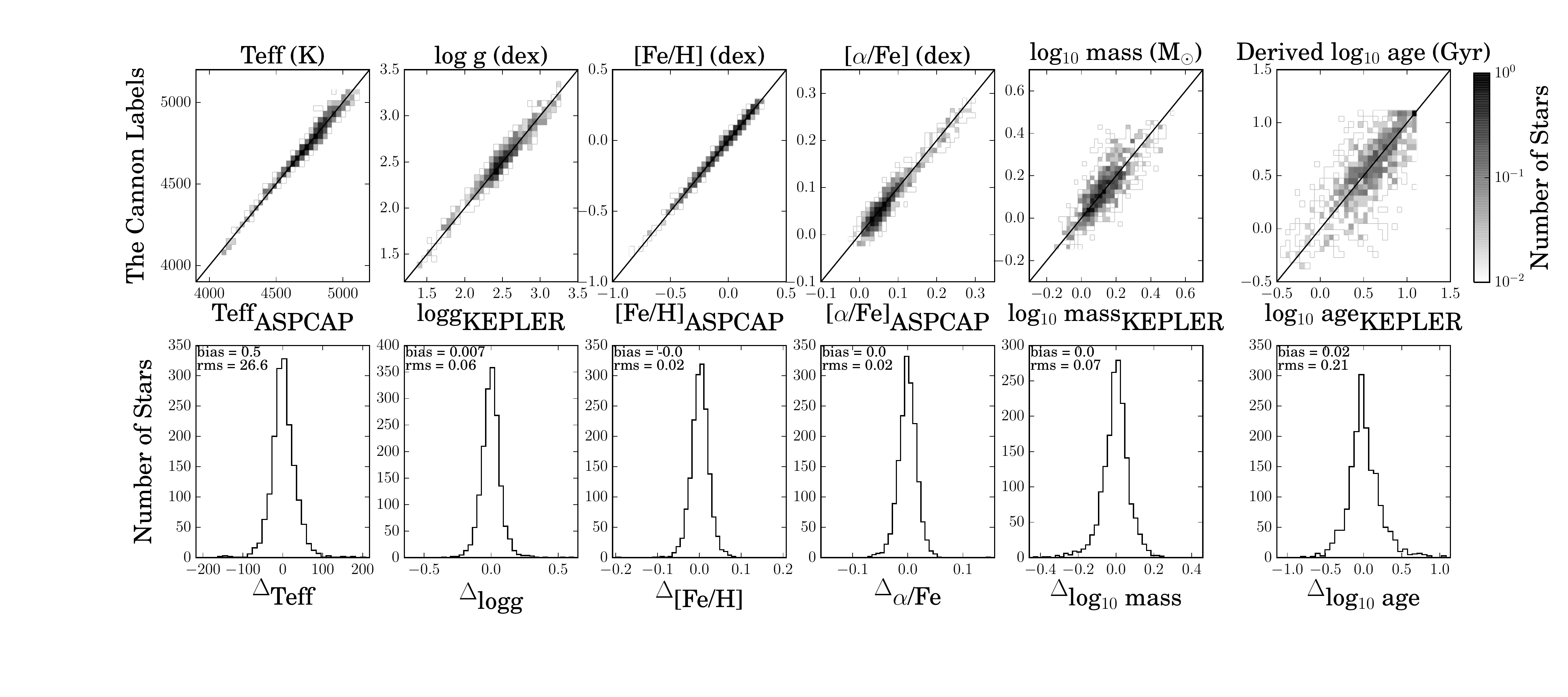}
  \caption{Cross validation of the training dataset of 1639 stars, for the \teff, \logg, \feh, \alphafe\ and mass labels: the results for \tc's labels for training performed on 90\% of the \apokasc\ stars, showing the performance at test time on the 10\% of the stars not included in training, run 10 times. The panel on the far right is the derived age label from the mass determined with \tc, using interpolation with PARSEC isochrones. The 31 stars with a $\chi_{reduced}^2$ statistic of $>$ 2 (2\% of the training data) have been removed.}
\label{fig:validation1}
\end{figure}

To determine the uncertainties from \tc\ on our individual \teff, \logg, \feh, \alphafe\ and mass measurements, we perform a take-stars out test on the set of reference objects.
For the take-stars out test we train the spectral model iteratively on 90\,percent of the reference spectra and then run the test step on the remaining 10\,percent of the spectra and we do this 10 times, stepping through each next 10\,percent of the data. Our results are shown in \figurename~\ref{fig:validation1} for the five labels. The top panels show the cross validation results comparing the input and output labels and the bottom panels show the histograms of the $\Delta$(input - output) for each label. The training labels (x-axis in the top panel of the Figure) are from \aspcap\ and astroseismology, as described in Section 2 and the output labels (y-axis in the top panel) are from \tc.  The sixth panel in this Figure shows the masses transformed to ages using interpolation between the PARSEC isochrones, where the red clump evolutionary state has been adopted on the isochrones at each age and \feh. We have removed spectra that are not able to be well fit by \tc's model; where the reduced $\chi_{reduced}^2$ of the model from model fit to the data is $\chi_{reduced}^2$ $>$ 2, which corresponds to 31 stars removed from the 1639 sample.

This \figurename\ shows that \tc 's purely mathematical approach of label transfer estimates the stellar labels with accuracies of 31K in \teff, 0.07 dex in \logg, 0.02 in \feh, 0.02 in \alphafe\ and  0.07 dex in log mass, or 0.21 dex in the inferred log age (Gyr) over the label-range of the reference stars. Notably, the uncertainty on the mass (20 percent) is only slightly larger than the \apokasc\ catalogue uncertainty of 12\%. It is important to remember that the objects plotted are the left-out objects and the spectra of these objects are completely detached from the training step,  except that they have the same experimental set-up and are drawn from a part of label space that is represented by the remaining reference objects.

\subsection{\tc's generative model at the test step: the red clump stars}

\begin{figure}[hp]
\centering
      \includegraphics[scale=0.5]{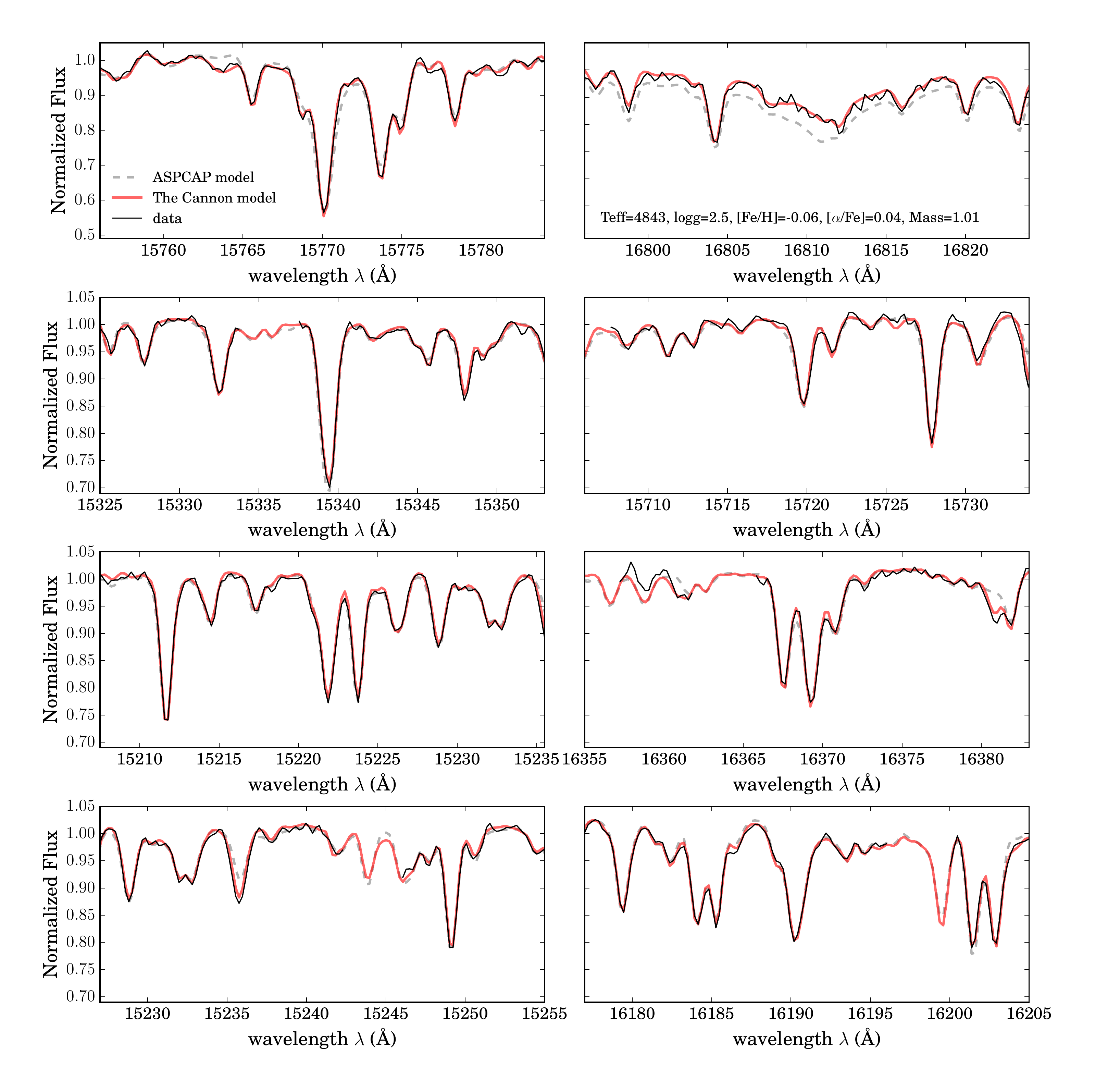}
  \caption{A spectrum of one of the red clump stars in the catalogue of \citet{Bovy2014} shown in black, with the best fit model from \tc\ in red. The eight wavelength regions shown correspond to the highest first-order coefficient amplitudes shown in Figures 2 to 5 (from top to bottom). The best fit model from \aspcap\ is shown in the grey dashed line.}
\label{fig:spectra}
\end{figure}

In Figure \ref{fig:spectra}, the spectrum of one of the red clump stars (not in the training set), which is representative of a typical red clump spectra in the \apogee\ red clump catalogue (discussed in Section 4.1), is shown along with the generative model from \tc\ at its stellar labels and the best fitting model from \aspcap. The data is shown in black, the synthesized model from \tc\ is shown in red and the best fit model from \aspcap\ is shown in the grey dashed line. The wavelength regions shown in this Figure are those for which the highest amplitude of the coefficients are located (Figures \ref{fig:logg} -- \ref{fig:mass}).  This example red clump star has parameters of \teff\ = 4843 K, \logg\ = 2.5 dex, \feh\ = --0.06 dex, \alphafe\ = 0.04 and mass = 1.0, determined by \tc.

Figure \ref{fig:spectra} illustrates that the generated spectral model from \tc\ provides a very good fit to the survey spectra. In fact the generative model from \tc\ is a better fit to the data than the best fit synthetic model from \aspcap. That the model from \tc\ is a good fit to the data demonstrates that the five-labels that we use to train, as well as our polynomial model, are sufficient to very well describe the behavior of the flux of a typical red clump star, given the training set of reference stars from the \apokasc\ catalogue. Note in this Figure, that one of the regions where \aspcap\ performs most poorly is at the Brackett line (see Section 3.3.1), which is highly \logg\ sensitive (Figure \ref{fig:logg}). This may indicate a problem with the model stellar atmospheres or its associated oscillator strength for this feature (or the lines it is blended with).

\subsection{Which parts of the H-band spectra constrain the labels?}

\tc\ is a generative model that determines a coefficient at every pixel, or wavelength. These coefficients describe how the flux depends on the stellar labels, given the model (in this work, in Equation \ref{eq:quad}).   A near-zero coefficient for a given pixel indicates that the flux at that pixel is independent of the labels. Conversely, the largest values of the coefficients are where the spectra changes most significantly with the label or labels. Here we examine the origin of the highest coefficients for the first order, linear coefficients. We use the first order coefficients to identify some key regions of the spectra that contain the most information with respect to the labels, and to determine which elements and molecules in particular these coefficients correlate with. 

Figures \ref{fig:logg} to \ref{fig:mass} show the first order coefficients of our model described in Equation \ref{eq:quad} over a narrow wavelength range ($\approx$ 30 Angstroms), centered on where the first order coefficients reach their largest amplitude. The 0$^{\mbox{th}}$ order coefficient is shown in the top panel and the first order linear terms $\theta_l$, where $l$ = \teff, \logg, \feh, \alphafe\ and mass, are shown in the middle panels. For a given set of labels, we can use \tc's model to generate the spectra (using all coefficients). The generated spectra is shown in the bottom panel of each Figure, for a representative set of stellar labels. These spectra are made at three steps across each stellar label, for each respective first order coefficient. This directly illustrates how the flux changes with each label in regions where the coefficient associated with that label is highest.  

We use the DR12 \apogee\ linelist (Shetrone et al., 2015, submitted), Kurucz model atmospheres \citep{castelli2004} and the stellar synthesis code MOOG \citep{sneden1979}, to determine which elements correspond to the absorption features in the spectra where the highest first order coefficients are located. The elements and molecules are marked on the 0$^{\mbox{th}}$ order coefficient spectra in the top panel of each of the Figures. The 0$^{\mbox{th}}$ order-coefficient vector $\theta_0$, or the baseline spectrum of the model is, essentially, the intersect spectrum of the training set of stars. 

The absorption features in the H-band are heavily blended with OH, CN, CO and $^2$C molecules and the figures indicate which absorption features are comprised of blends of molecules and elements, at the stellar parameter space of \apogee\ stars. The elements that show the most significant changes with the labels show gratifying accord with expectations from stellar physics and these are discussed below, for each of the five labels. 

\subsubsection{Spectral dependencies on \logg} 

\begin{figure}[hp]
\centering
    \includegraphics[scale=0.51]{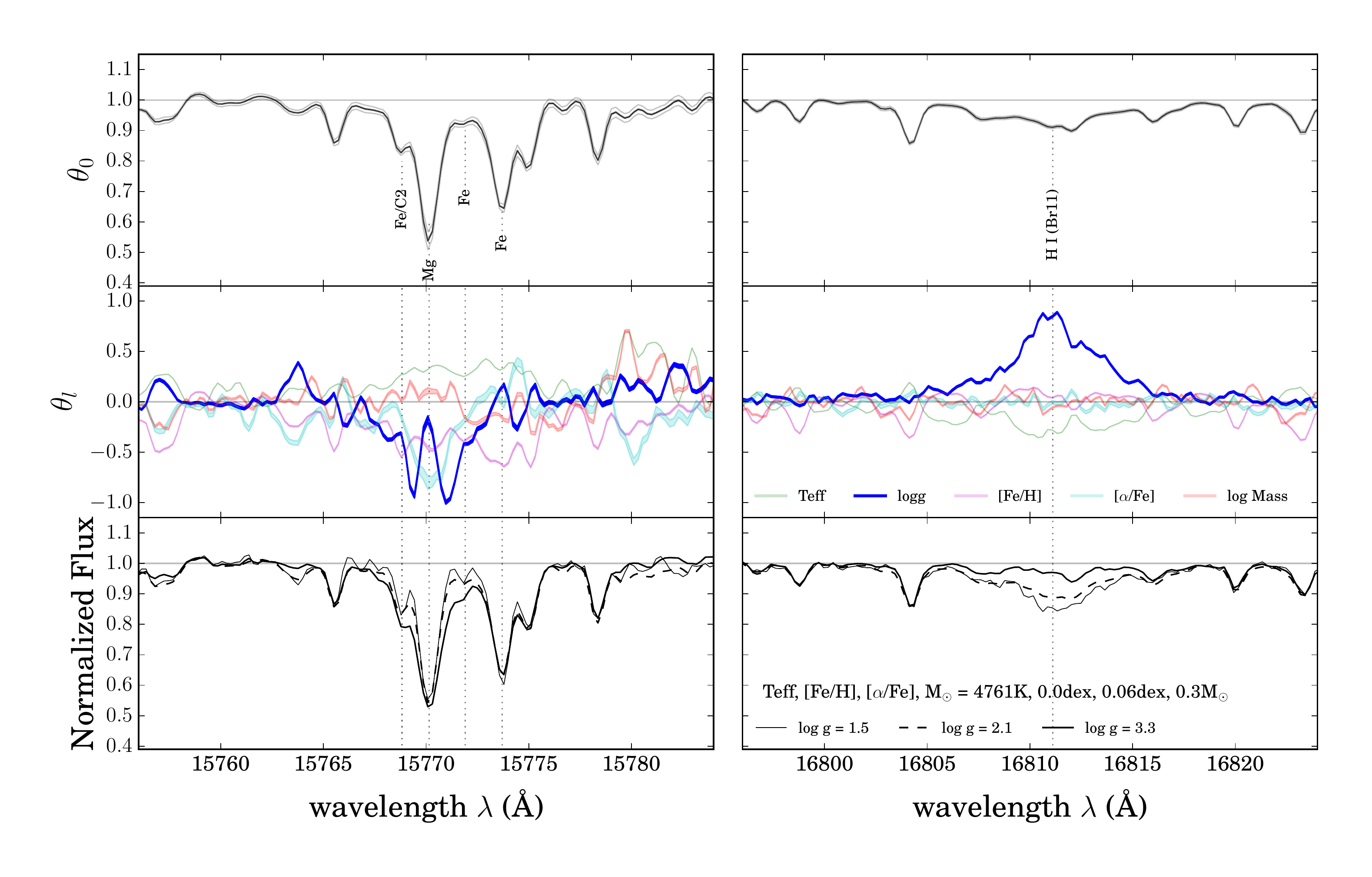}
  \caption{The 0$^{\mbox{th}}$ and first order coefficients ($\theta_0$ , $\theta_{\mbox{$l$}}$) of the model trained on the \aspcap\ stars showing the two 30 Angstrom wavelength regions where the \logg\ coefficient (in dark blue) reaches its highest absolute amplitudes. The 0$^{\mbox{th}}$ coefficient (at top) describes the intersect spectrum and a number of spectral absorption features are marked. The middle panel shows all coefficients, each normalized to their highest amplitude, for all \textit{$l$} = \teff, \logg, \feh, \alphafe\ and mass coefficients, where all coefficients are shown in transparent lines except for the \logg, which is of interest here. The bottom panels show the generated spectra from \tc's model, for three increasing values of \logg, which span the full range of \logg\ values in the training set. The selected parameters for this star are set around the fiducial point of the stars in the training set, and represent a typical \apokasc\ spectra. The panel at left where the \logg\ coefficient reaches its highest amplitude shows the \logg\ information is concentrated to the wings of the strong Mg feature in the spectra, dissimilarly to the \alphafe\ information which is in the core as seen from the cyan coefficient. The panels at right show the broadening of the flux as a consequence of the H Brackett-11 line which, like Mg,  is similarly gravity sensitive due to pressure broadening effects (see the text). The labels of the fiducial spectra are indicated in the bottom right hand panel.}
\label{fig:logg}
\end{figure}

Figure \ref{fig:logg} shows two 30 Angstrom regions of the spectra centered on the two highest first-order \logg\ coefficient amplitudes, $\theta_l$ = $\theta_{logg}$. The three panels at left show the highest \logg\ coefficient and the three panels at right show the second highest coefficient. Relevant elements and molecules that correspond to the absorption features are marked at top on the baseline spectrum of the model.  The middle panel of Figure \ref{fig:logg} shows the first order coefficients $\theta_l$ that are linear in \teff, \logg, \feh, \alphafe\ and mass. The coefficients have all been normalized to their largest absolute value, so that an amplitude of $\theta_l$ = 1 for any coefficient is at the highest value.  

The bottom panel of Figure  \ref{fig:logg} shows a generated spectrum from \tc's model for a reference set of stellar parameters, which are the mean of the training set labels, or the fiducial spectra. These reference labels are set at \teff = 4761K, \feh = 0.0 and \alphafe = 0.06, mass = 0.3 $M_{\mbox{star}}$, for three different \logg\ values of \logg\ = 1.5, \logg\ = 2.1 and \logg\ = 3.3. From the centre left panel of Figure \ref{fig:logg} it is clear that the flux at any given pixel can correlate with multiple labels. Typically some coefficients, like \teff\ and \logg\ show an inverse relationship between label amplitude and flux.

The location of the highest amplitude \logg\ linear coefficient, which is shown in the top left-hand panel of Figure \ref{fig:logg}, corresponds to a strong  Mg feature in the \apogee\ spectra. Importantly, the highest amplitude of this coefficient corresponds not to the core of the Mg feature, but to the wings, and more strongly so for the upper wavelength side of the feature. The core of the Mg feature in fact  corresponds to a significantly lesser amplitude of  the coefficient; clearly in the case of \logg,  there is a dramatic reduction in the information content of these pixels in the core of the feature. Note where the \logg\ coefficient decreases from the wings to the core (in blue) the \alphafe\ label (in cyan) increases so that the largest amount of information in this region for the \alphafe\ label is, conversely, from the core of this feature. This Mg feature at 15770.15 Angstroms, is one of the two strongest Mg features (along with the Mg feature at 15753.29 Angstroms) across the \apogee\ H-band spectral region. 

That the strongest coefficient in \logg\ comes from the wings of a strong Mg line in the H-band \apogee\ spectral region is well-aligned with empirical analyses in other, more comprehensively studied wavelength regions. The wings of strong lines are known \logg\ indicators \citep{Gray2008}. Specifically, the wings of Mg lines in the optical wavelength region, which are sensitive to pressure broadening, are used by \citet{F1997} to derive \logg\ for F and G main sequence stars.  

Similarly, Brackett lines (as well as Balmer and Paschen lines)  are sensitive to pressure (Stark) broadening and are therefore excellent tracers of \logg\ in stars. The second highest amplitude coefficient for the first order linear \logg\ coefficient in the \apogee\ spectral region is at the Brackett feature  at $\approx$ 16810 Angstroms, as shown in the right hand panels of Figure \ref{fig:logg}. The  bottom panel of this Figure (at right) shows how significantly the flux varies as a function of \logg\ for this feature. In addition to being second highest in amplitude, the sign of this coefficient for this feature is positive and opposite to that  of the wings of the Mg line. As seen in the bottom panel at left, the wings of the Mg feature deepen with increasing \logg, where as for the Brackett feature at right, the spectral profile flattens with increasing \logg, for any given set of stellar \teff, \feh, \alphafe\ and mass parameters, directly demonstrative of the inverse relative relationship between the two features.

\subsubsection{Spectral dependencies on \teff} 

\begin{figure}[hp]
\centering
    \includegraphics[scale=0.51]{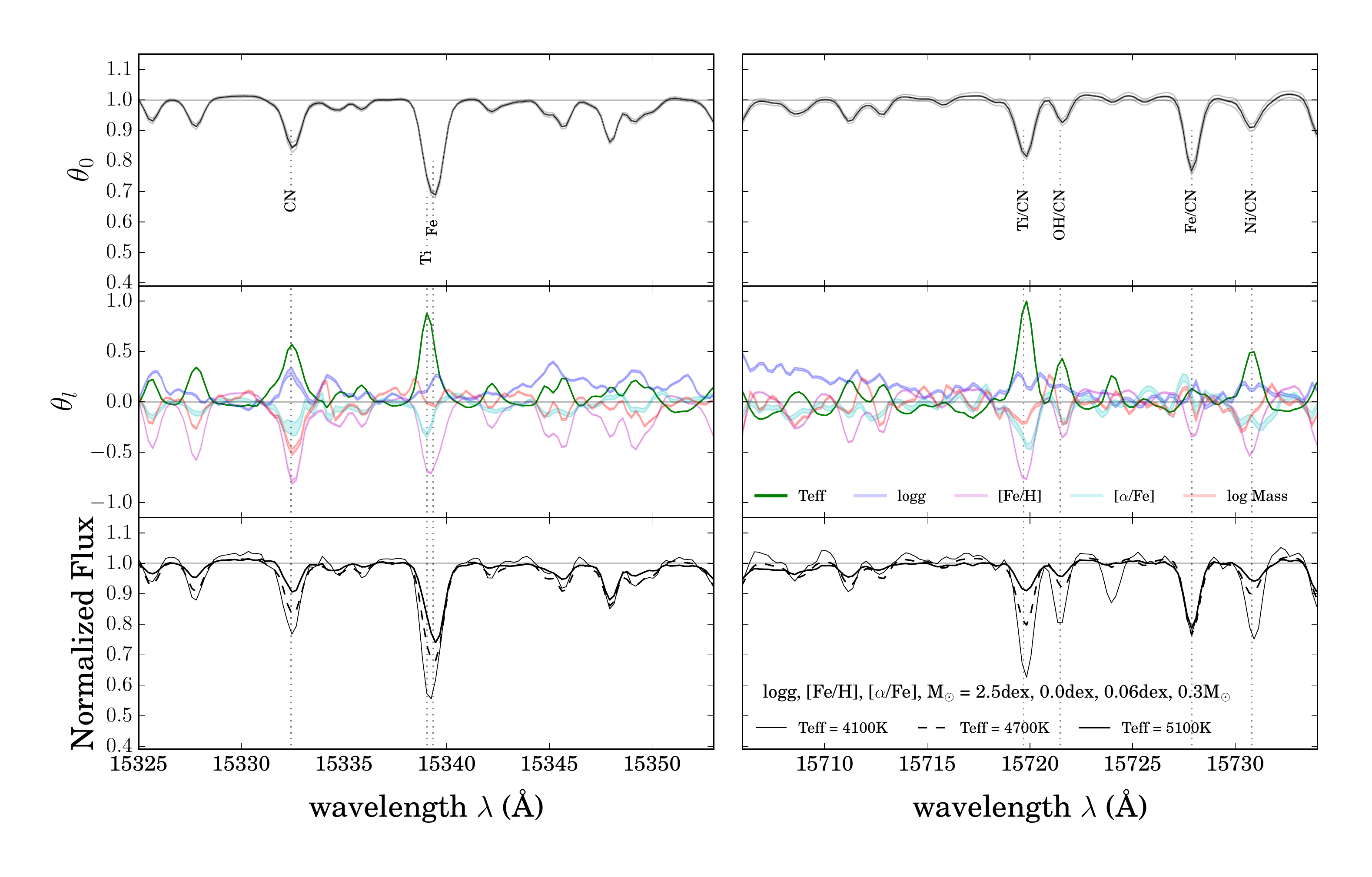}
  \caption{Same as for Figure \ref{fig:logg} but showing the two 30 Angstrom regions centered at the highest \teff\ coefficient (in green). The \teff\ coefficient is typically anti-correlated with the \feh\ and \alphafe\ coefficients and traces the absorption profiles, reaching the highest amplitudes at the core of metal lines across the entire spectrum.}
\label{fig:teff}
\end{figure}

Figure \ref{fig:teff} shows the same information as for Figure \ref{fig:logg} but for the two highest \teff\ coefficients, $\theta_l$ = $\theta_{\mbox{Teff}}$, centered on $\approx$ 15338 and 15720 Angstroms. The highest \teff\ coefficients correspond to the cores of two Ti lines in the H-band spectra (one of which is blended also with Fe and the other with CN). The temperature coefficient is typically positive in the \apogee\ spectral region, with exceptions, for example at the Brackett feature shown in Figure \ref{fig:logg} (where it is inversely correlated with the \logg\ coefficient). \teff\ is typically strongly anti-correlated with \feh\ and \alphafe, as seen in Figure \ref{fig:teff}. This anticorrelation reflects that in a spectrum at a given \feh, as the temperature increases the lines weaken and so the flux decreases, whereas at a given \teff, as the metallicity increases the lines strengthen, and the flux increases. 

As we have a coefficient at \textit{every} wavelength, which we can map to the chemical elements and molecules in the spectra using the \apogee\ linelist, we can interpret the spectral relationship between labels and flux in more detail than for an integrated absorption feature itself.  For example, there is an asymmetry in the variation of the \teff\ label in the left-hand bottom panel of Figure \ref{fig:teff}. This asymmetry likely reflects the changing ratio of the blends within this absorption feature (in this case, the feature is a blend of Ti and Fe, which are offset within this feature in their central wavelength).

The coupling of the data-driven model to stellar physics and mapping to the elements or molecules that determine the flux has important applications for stellar astrophysics. Here our aim is simply to verify that the information in the spectra or regions of highest spectral dependence on the labels, originates from genuinely sensible and plausible chemical features in spectral space.

\subsubsection{Spectral dependencies on \feh\ and \alphafe} 

\begin{figure}[hp]
\centering
    \includegraphics[scale=0.51]{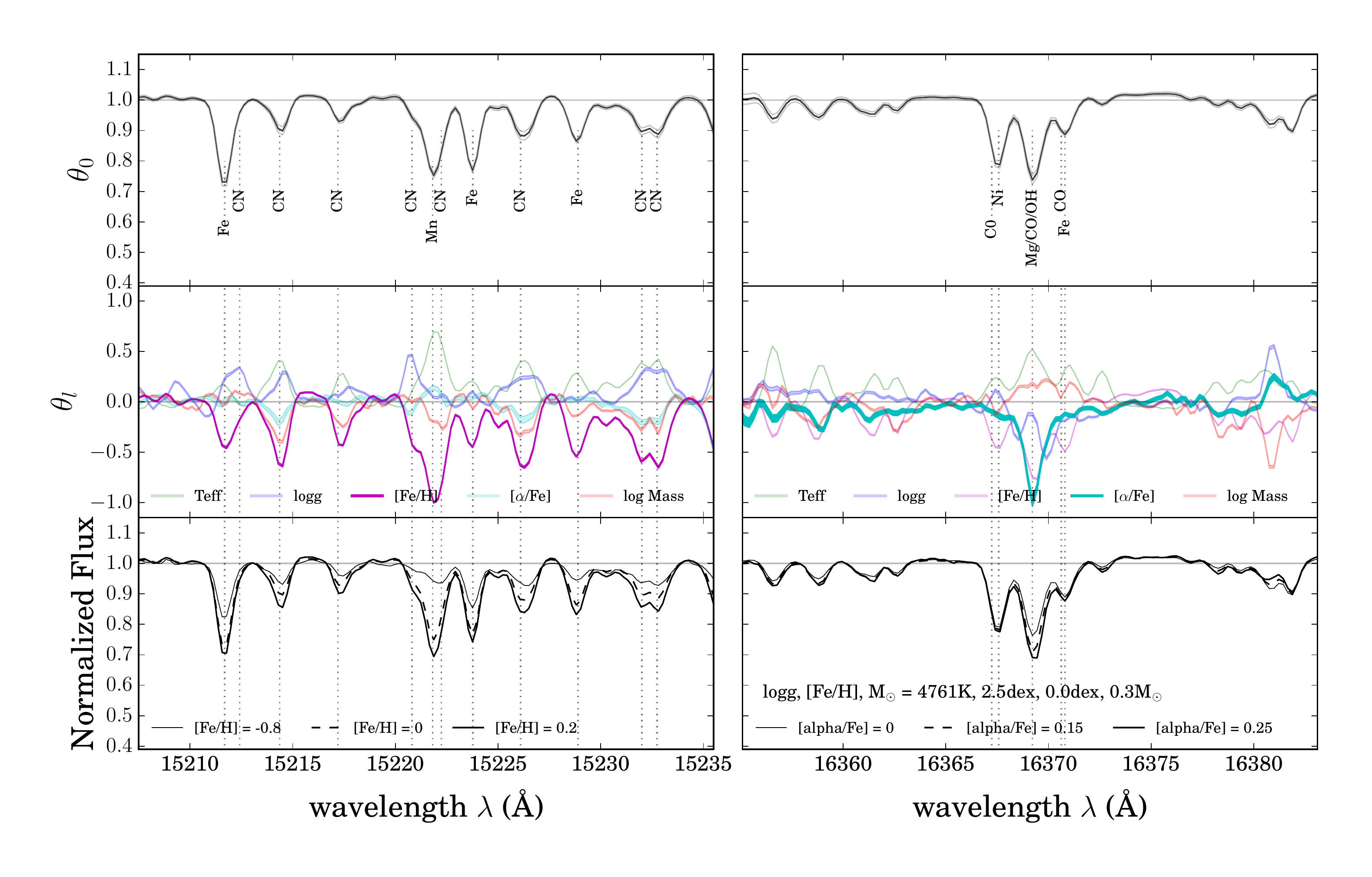}
  \caption{Same as for Figure \ref{fig:logg} but showing the two 30 Angstrom regions centered at the highest \feh\ coefficient (in magenta), at left and highest \alphafe\ coefficient (in cyan), at right. The largest coefficient in \feh\ is from an Mn (iron-peak) line which correlates with \feh. Note that for this line in the mid-panel at left, the \alphafe\ coefficient is near-zero where the \feh\ information is highest. The highest amplitude of the \alphafe\ coefficient is seen in the core of one of the strong Mg lines in the \apogee\ spectra, shown in the panel at right.}
\label{fig:feha}
\end{figure}

Figure \ref{fig:feha} is demonstrative of the highest \feh\ and \alphafe\ coefficient in the spectra, $\theta_{[Fe/H]}$, $\theta_{[\alpha/Fe]}$. The \feh\ and \alphafe\ labels are typically correlated with the cores of all of the absorption features in the spectra, particularly for the \feh\ label, as seen at left. This is unsurprising as the overall metallicity,  [M/H] of  a star simply correlates with the \feh\ and the \alphafe\ is known to increase with \feh\ and flattens to a plateau at high \alphafe\ and low \feh, subject to the star formation rate and initial mass function. For many (but not all) absorption features, the \teff\ shows an inverse correlation with temperature as seen in the left hand panel of Figure \ref{fig:teff}. 

The strongest \feh\ coefficient corresponds to a core of a (blended) Mn feature, the flux of which changes dramatically as a function of \feh\ over the range of  --0.8 $<$ \feh\ $<$ +0.2 as shown in the bottom panel of Figure \ref{fig:teff}. Mn is one of the Fe-group elements (in addition to V, Ti, Cr, Co and Ni) and this element is known to correlate directly with \feh\ \citep[see][]{Maria2008, B2015}. 

The largest coefficient in \alphafe\ corresponds to the core of the strong (alpha-element) Mg line at $\approx$ 16370 Angstroms (which is also blended with CO and OH) and dissimilarly to the \logg\ coefficient, it is the core of the line that correlates with \alphafe. Note that for the \logg\ coefficient at this blended Mg feature in the middle panel of Figure \ref{fig:feha} (right), the \logg\ coefficient is $\approx$ 0 at the very centre of the line profile. The log g coefficient increases to a a much larger amplitude in the wings of the feature.

\subsubsection{Spectral dependencies on mass} 

Finally, having verified that \tc\ delivers physically sensible origins of the \teff, \logg, \feh\ and \alphafe\ labels, we examine the origin of the mass information from which we can infer age of \apogee\ stars. 
These first four labels are most straightforward  to convincingly derive: indeed they are standard labels that are routinely determined from a stellar spectrum. However, delivering a mass label, mass, directly from stellar spectra marks a significant step forward in the exploitation of stellar spectra. 

With the exception of a few specific indices which have been used previously to derive mass and inferred age, this work is the first claim of the success of a generalized approach for the extraction of stellar mass from spectra.  Mathematically this works (as per the cross validation) and now we examine the interpretability of the mass label in terms of direct spectral signatures.  For example, consider the case of main sequence stars.  In this case, mass is correlated strongly with effective temperature and more weakly with surface gravity in a composition dependent fashion.  A Cannon-like approach with mass labels would therefore be likely to have very similar spectral correlations, producing something equivalent to isochrone fitting for photometry.  Red giants have a wide range in log g and different mass tracks differ only subtly in effective temperature, with a very strong metallicity dependence.  One might therefore fear that any mass estimates would have very large random uncertainties, which is clearly not the case based on our results from Section 3.1.

From previous analyses in the UV and optical wavelength regions, we might expect spectral mass indicators, if present, to be realized in (i) chromospheric activity (emission), (ii) dredge-up effects (and changing line strengths and profiles of particular elements) or (iii) some combination of individual elemental abundances that reflect the enrichment history of the Milky Way with time (changing element ratios in the spectra). 

\begin{figure}[hp]
\centering
    \includegraphics[scale=0.51]{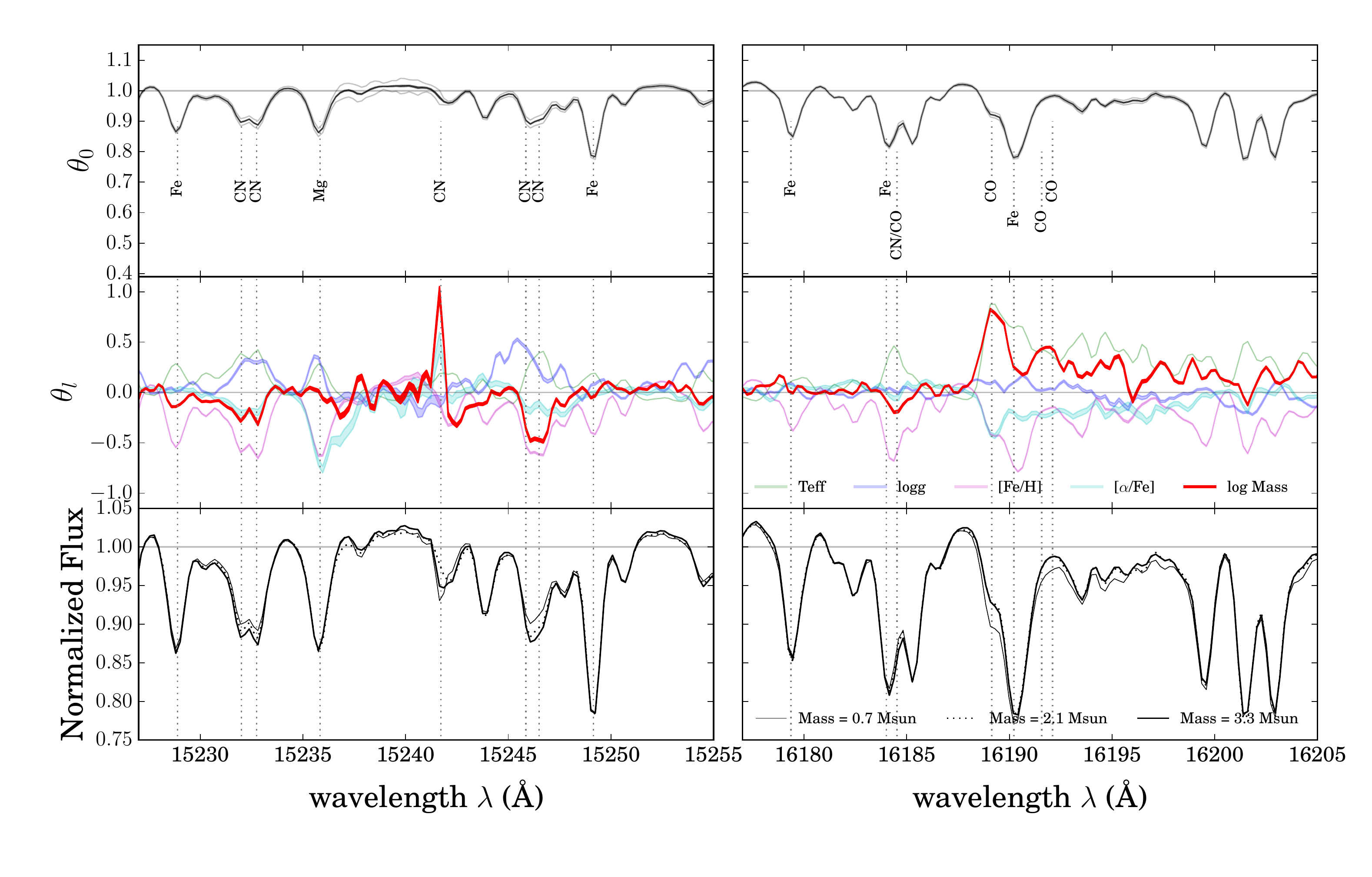}
  \caption{Same as for Figure \ref{fig:logg} but showing the two 30 Angstrom regions centered at the highest mass coefficients (in red). The bottom panels show that that spectra change line shape and depth with varying mass and the mass coefficients are highest in region of CN and CO molecular blends. In particular, these blends corresponding to the highest mass coefficient in this Figure comprise both $^{12}$C and $^{13}$C. Where the mass coefficients are negative, the flux is shallower at lower mass. Where the mass coefficient is positive, the flux is deeper at lower mass.}
\label{fig:mass}
\end{figure}

Figure \ref{fig:mass} shows the two largest coefficients in the log mass label. The information for the mass label is from (the relatively weak) CN and CO molecular features. Although we show only two regions as demonstrative, we have verified that the five highest mass coefficient amplitudes all correspond uniquely to predominantly CN but also CO molecular features. The relationship between mass and CN is consistent with the discovery by Martig et al., (2015)., which shows that the [C/N] ratio calculated from \apogee's delivered catalog of C and N abundances in data release DR12 correlates with the mass and inferred age of the \apokasc\ stars. \citet{Sal2015} also demonstrate the theoretical basis for the [C/N] ratio as an age indicator from the post first dredge-up. Martig et al., (2015)., use the C and N abundances to create a model from these abundances and known masses of the \apokasc\ stars. With \tc\, this information is similarly exploited, only at the spectral level: we do not inform \tc's generative model about the origin of the information (instead we rely on stellar physics to interpret the regions where the information is highest). 

From the synthesized spectra in the bottom left-hand panel, it is apparent that it is not only the line strength that changes with the mass coefficient, but also the line profile. Furthermore, the mass coefficient correlates with the \feh\ coefficient at the regions of the CN blends, at left and anti-correlates with the \feh\ coefficient at the CO molecular feature, at right. Where the coefficient is positive, the flux of the model becomes larger at lower mass where as when the coefficient is negative, the line strength is weaker at smaller mass.  

The changes in spectra as a function of mass are in general very subtle compared to the other labels. This is likely responsible for the relatively large scatter in the mass label determined with the take-stars out test, shown in Figure 1. Note the other stellar labels have been historically well determined from spectra, even without general mathematical methods like \tc\ (which can optimally exploit all of the available and potential information). The correlations contained in the traditional stellar labels of \teff, \logg, \feh\ and \alphafe\ are more straightforward to extract (e.g. \feh\ correlates with the cores of most absorption features in the spectra). This highlights the strength of an approach like \tc\ to determine and quantify the information that can be truly extracted from data, particularly as a function of signal to noise. 

Examining the CN molecular regions in more detail, the two CN regions shown in Figure \ref{fig:mass} with the highest mass coefficients are in fact a blend of CN molecules containing both $^{12}$C and $^{13}$C. Similarly, the CO feature at right is a blend of both $^{12}$C and $^{13}$C. It is this ratio which may drive the changing line profile as a function of mass and may play an important role in delivering the mass information from \apogee\ spectra. This is because the $^{12}$C/$^{13}$C ratio is known as one of the best diagnostics of deep mixing in stellar interiors and so is known to contain information with respect to stellar mass.  

Changes in isotope ratios complement information from the carbon to nitrogen ratio, and the combination is more powerful than any one indicator.   In addition to their diagnostic power for mass, they also serve as markers of evolutionary state; carbon isotope ratios have already been used in the literature to differentiate first ascent giants from red clump stars \citep[][and references therein]{Taut2013}.  The empirical data therefore naturally accounts for both the traditional first dredge-up effect, incorporating material processed in the core of the main sequence precursor \citep{Taut2010}, and in situ giant branch mixing \citep{Gilroy1991}.  This is true even in the absence of a predictive theory for the origin of the latter phenomenon.

\subsection{Mass (and age) determination at a given \feh\ and \alphafe}

We use small regions of mono-abundance space, that is, regions of a small range in \feh\ and \alphafe, to demonstrate that we have a bona fide spectral mass label, from which we infer stellar ages. We can thereby show that our mass/age label we does not reflect simply some combination of the other four labels. We wish to illustrate in particular, that the mass label that we use to infer age is not simply another expression of the \alphafe\ label. The \alphafe\ label itself is often used as an overall age proxy in abundance studies given gross expectations from stellar evolution and chemical yields in stellar populations. 

\begin{figure}[hp]
\centering
  \includegraphics[scale=0.4]{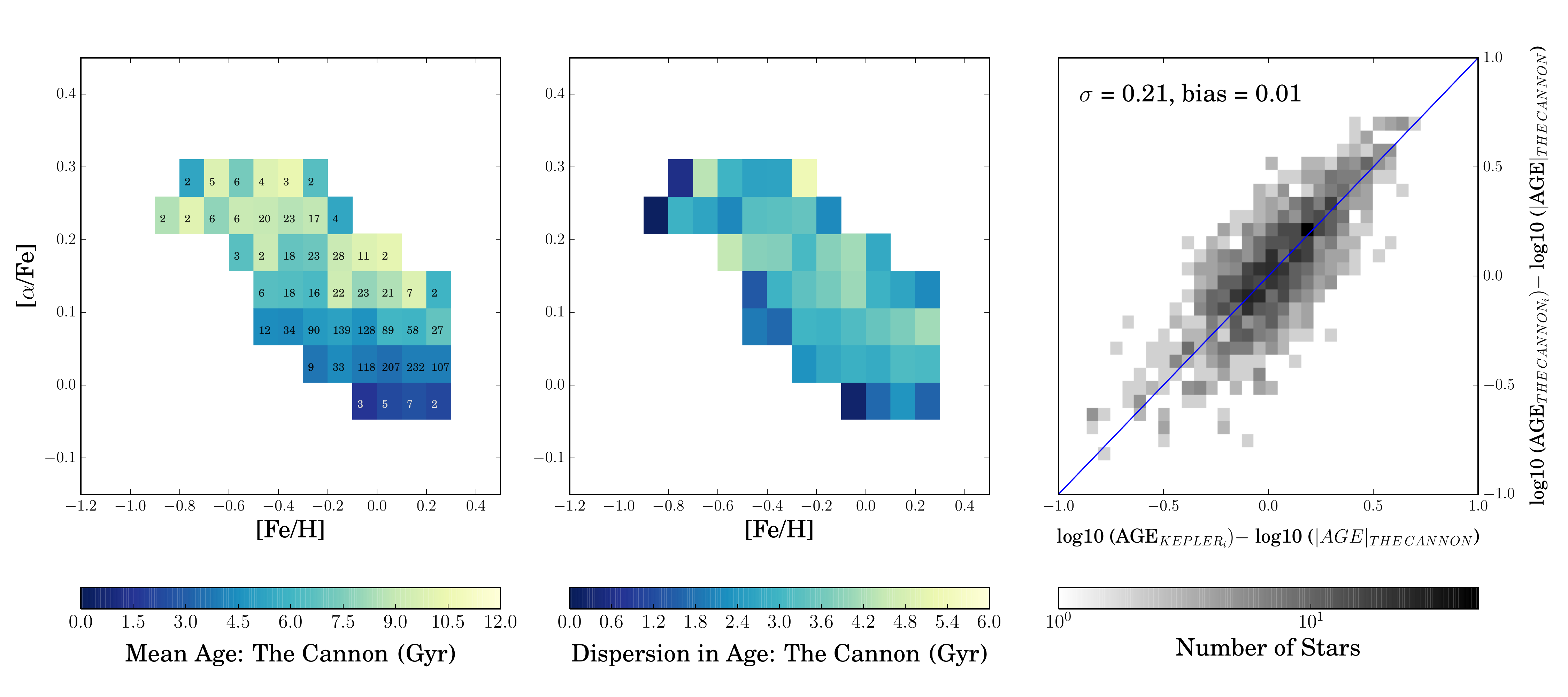}
    \caption{The \feh-\alphafe\ planes at left and centre for the training set of \apokasc\ stars, colored by mean age of the stars in each bin at left,  and the age dispersion at centre, where the number of stars in each bin is indicated at left. The far right hand panel shows the stars in the \feh-\alphafe\ plane where the individual stars $i$ have been subtracted from the mean age in their respective bin, taking both the age from \tc\ and the age from \kepler. The panel on the far right shows the correlation in the input and output age label delivered by \tc\ in cross-validation within a narrow bin in abundance space.  }
\label{fig:alphabins}
\end{figure}

In Figure \ref{fig:alphabins}, the \apokasc\ set of reference stars used to train \tc\ are shown in the \alphafe-\feh\ plane. These stars are binned into small mono-abundance boxes in this Figure and the panel at the far left indicates how many stars are in each of these bins. The colormap represents the mean age and the age dispersion from \tc\ that is obtained in cross-validation.

The input label for the inferred age is from the seismic scaling relations for these objects (from \kepler) and the output label is derived from the inferred age from the mass label output by \tc\ in cross validation (the take star out test in Section 2). The far right hand panel of Figure \ref{fig:alphabins} shows the individual age label for each star, from \tc\ (on the x-axis) and from \kepler\ (on the y-axis), subtracted from the mean age value in each age mono-abundance bin. This is done for each bin and combined in this right hand panel in the Figure. If there was no additional information in each of the mono-abundance bins with respect to age, that is, if the age information was simply a reincarnation of the \alphafe\ label then there would be no expected correlation between the difference in \tc\ and \kepler\ and the mean age. That there is a 1:1 relation between these two axes reflects that \tc\ works mathematically to determine the mass label and that the mass label within a mono-abundance bin carries additional information.

\section{Masses and Ages for APOGEE Red Giant Stars in DR12} 

For the following sections, the results for the ages of stars are inferred from their output mass label determined by \tc. We train on log mass, as described in Section 2.2, where the mass label for the \apokasc\ stars has been determined using the standard seismic scaling relations. We transform the output mass to age as described in Section 2.3, for mapping the age distribution of the red giant stars in \apogee's DR12 across the disk, as shown in Figures \ref{fig:redclump_age} and \ref{fig:allage}. Our stellar labels determined by \tc\ for 80,000 red giant stars from DR12  (including $\approx$ 20,000 red clump stars) are provided in an online Table, with a partial extract shown in the Appendix in Table \ref{tab:online}.

\subsection{Stellar ages for the red clump sample}

The \apogee\ DR12 sample comprises primarily red giant stars plus a valuable subset of  $\approx$ 20,000 red clump stars, identified by \citet{Bovy2014}.  These stars have individual distance uncertainties of 5\,percent. These red clump stars cover a large radial extent of the disk, spanning distances of 4 -- 15 kpc and are located predominantly at heights $|z|$ $<$ 3.0 kpc from the plane. The red clump sample is a representative and unbiased sample of Milky Way disk stars and has an expected age distribution peaking at about 1.8 Gyr with a tail out to old ages \citep[see Figure 15 of][]{Bovy2014}. We take our model, trained using the reference \apokasc\ stars and determine the stellar parameters and masses for these red clump stars. We then infer ages by interpolating in label space onto PARSEC isochrones.

The red clump may seem to be a surprising choice to use for age studies, as stars in this evolutionary state are known to experience stochastic and significant mass loss relative to prior epochs.  However, we do account for this mass loss, and given other uncertainties, age estimates for red clump stars are not dramatically more unreliable than those for first ascent giant branch stars \citep[see][for a recent discussion]{Casagrande2015}. The higher age uncertainties are also compensated for to some degree by having more reliable distances.

\subsubsection{The stellar age distribution of the Milky Way's disk across 4 --15 kpc}

We have determined the masses and (from PARSEC isochrones) inferred the ages for the $\approx$ 20,000 red clump stars that have distances known to approximately 5 percent. We use these results to show the age distribution of the Milky Way's disk. The full catalogue of the stellar labels determined with \tc\ for the red clump sample is included in Table \ref{tab:online}, in the Appendix. This data represents the largest homogeneous sample of stars in the Milky Way with mass and associated age labels, and extends the age mapping of the Milky Way from the previous local neighborhood only (GCS) to trace the inner to outer disk, from 4 -- 15 kpc.

\begin{figure}[hp]
\centering
 \includegraphics[scale=0.5]{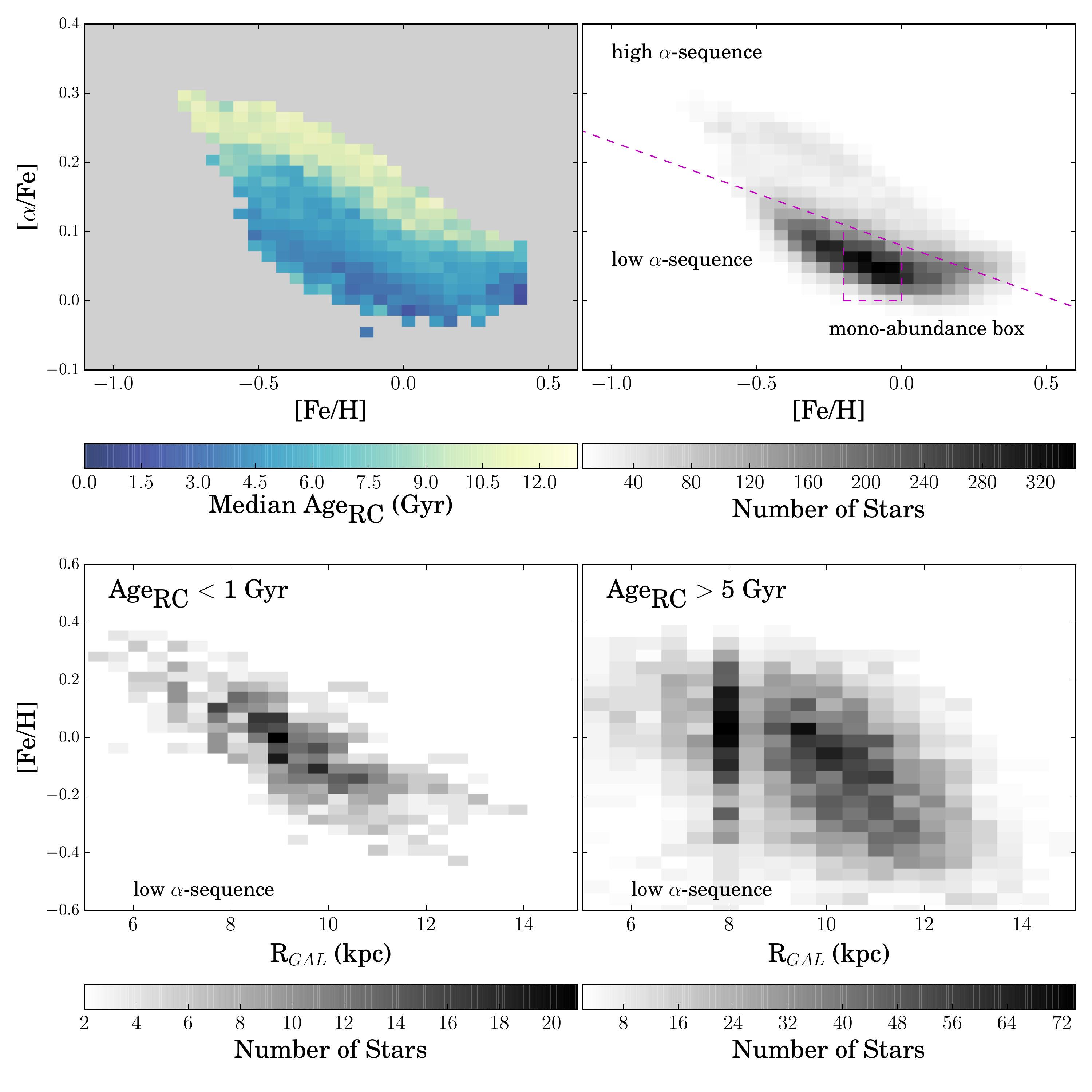}\\
    \caption{The $\approx$ 17,065 red clump sample of stars. The top left panel shows all stars colored by median age, where the median age$_{\mbox{RC}}$ does not represent the median age of the population from which it was drawn. The density distribution of this sample is shown at the top right and the sequences we use for examining the age of the disk, of the low$-\alpha$ sequence and mono-abundance population bins are indicated. The bottom panels show the \feh-\rgal\ distribution for the young and intermediate stars, in the low$-\alpha$ sequence.}
\label{fig:redclump}
\end{figure}

Figure \ref{fig:redclump} shows the median age of the red clump stars in the \alphafe-\feh\ plane (top left) and the density distribution of these stars (top right). We include the 17,065 stars with a $\chi_{reduced}^2$ of $<$ 2 in this figure, which excludes 15\% of the sample (all stars with their corresponding $\chi_{reduced}^2$ statistic are given in Table 1). Most of the red clump stars are located in the low-alpha sequence. We select the low-alpha sequence stars to examine the trends of the metallicity, \feh, as a function of radius, for the low compared to the intermediate age stars. This selection is for all stars below the dashed line in the density distribution of the clump stars, at the top right of Figure \ref{fig:redclump}. In the selection of these low-alpha stars we are selecting stars that should represent a single sequence of chemical enrichment. We therefore might expect differences in the distribution of \feh\ with radius for this sequence, as a function of age. This difference would be not due to different formation histories, but due to Galaxy evolution processes for this population, over time \citep[e.g.][]{Roskar2008, Schonrich2009}. 

The bottom panels of Figure \ref{fig:redclump} show density maps of the \feh\ of the youngest stars (at left) and the intermediate-oldest stars (at right), as a function of radius. At bottom left, there are 1669 stars with ages $<$ 1 Gyr and at right, there are 6716 stars with ages $>$ 5 Gyr. Note that there is an apparent over-density at about 8 kpc across all \feh, for the intermediate age selection. These are the stars in the \kepler\ field in the sample.

Importantly, the median age of the red clump sample is not the median age of the population from which is is drawn. The red clump age distribution, from stellar evolution theory, is peaked at young ages. As discussed in \citet{Bovy2014}, the red clump population is a long lived evolutionary phase (and one for which precise distances can be determined), and is an excellent population tracer. At the same time, the fraction of the mass in the red clump is a function of the overall star formation history or age distribution of the Milky Way's disk.  The red clump, whilst being an excellent tracer of the Milky Way disk, does not represent the unbiased stellar distribution function of ages in the Milky Way disk. 

The \feh\ distribution for intermediate age stars, as a function of \rgal\ is less tightly correlated with radius compared to the youngest stars in the red clump sample. That the \feh-radius correlation weakens with age likely reflects dynamical evolution processes in the Milky Way which redistribute the stars in the disk, such as radial migration. Intermediate stars, being longer lived, would have experienced a more significant dynamical timescale over which these processes take effect and so are scattered more from their original birth radii.  The youngest stars have been subject to a shorter dynamical evolution history and their current origin likely more tightly traces the origin of their birthplace, reflected in the correlation between radius and \feh, tracing the chemical enrichment of the gas which increases toward the centre of the Galaxy.  

The top right hand panel of Figure \ref{fig:redclump} shows a small box in the \feh-\alphafe\ plane from which we select stars for conditioning our age analysis on abundances. We use this mono-abundance box to investigate and compare the mean trends of age across the disk $(R_{GAL},z)$, contrasted with that for all stars, in demonstrating the information in the age label, even conditioned on abundances.

\begin{figure}[hp]
\centering
        \includegraphics[scale=0.43]{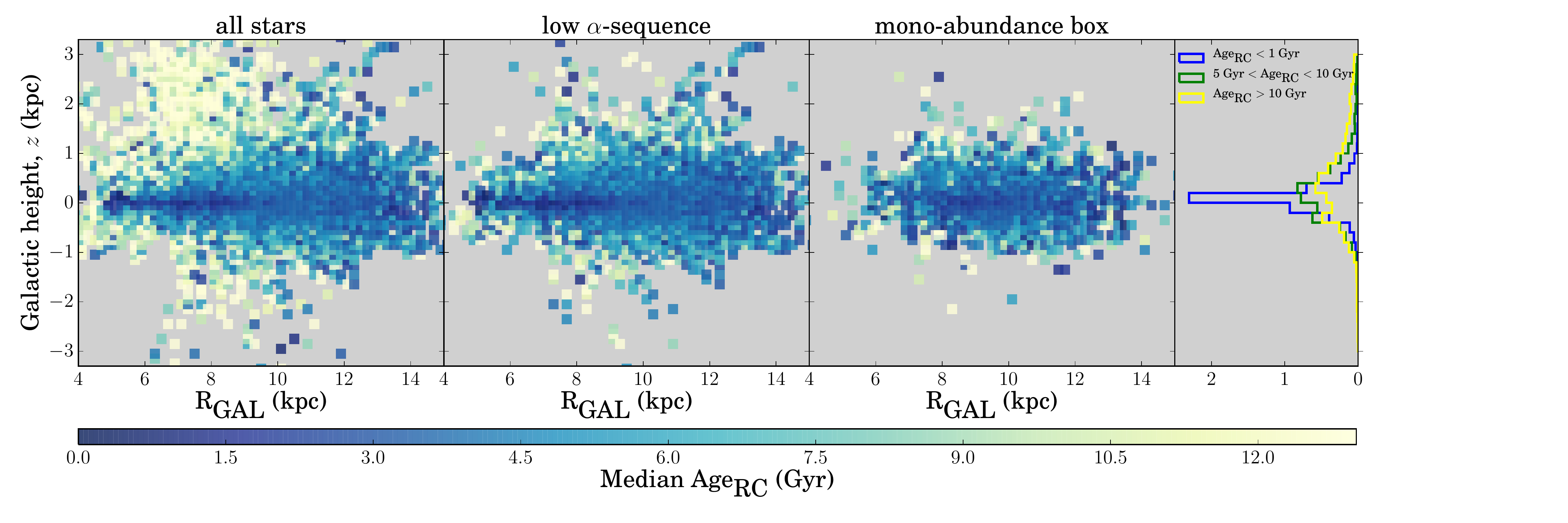}
    \caption{$(R_{GAL},z)$ Maps of the median age of the  \apogee\ red clump stars showing $\approx$ 17,065 stars at far left, the low-alpha-sequence only in the second panel from left and a small abundance bin in the low-alpha sequence third panel from left. The final panel, at right is a histogram of the different age distributions as a function of age, showing all stars across $z$, for a young, intermediate and old selection.  }
\label{fig:redclump_age}
\end{figure}

Figure \ref{fig:redclump_age} shows the $(R_{GAL},z)$ distribution of the red clump sample colored by median age across 4 - 15 kpc for  (i) all stars, at left, (ii) the low-alpha sequence, second from left and (iii) the mono-abundance sample bin shown in the top right panel of \ref{fig:redclump}, third from left. The distribution of young, intermediate and old age stars, for all of the stars (far left panel) is shown in the histogram at far right. 

Figure \ref{fig:redclump_age} demonstrates that the stars in a narrow $|z|$ range in the plane are typically young, spanning the radial extent of the sample. There are fewer stars in the low-alpha sequence far from the plane and the low-alpha sequence is dominant at larger radii (e.g. Hayden et al., 2015), however the same trends are seen in all three panels of age distributions. Older stars are present, preferentially at smaller radii as seen most clearly in the far left panel, and these are typically located further from the plane than younger stars. Stars transition to older ages further from the plane as the radius increases and there is an apparent vertical flaring in the age distribution with radius, with younger stars also dominating the ages at larger heights from the plane at the largest radii. 

The histogram at far right shows the very different distributions of young and old stars. The number of youngest stars is strongly peaked near $z$ $\sim$ 0 as these stars are concentrated to the plane, suggesting ongoing star formation in the gas enriched regions of the Galaxy. The older stars show a much broader distribution and extend to larger heights from the plane and are present in larger relative fraction at smaller radii, preferentially at larger $z$.  The younger stars extending out to larger radii including farther from the plane supports an inside out formation scenario for the Milky Way. These distributions which show young stars also at large heights from the plane imply that younger stars are born also at relatively large heights from the plane.

The centre and far right panels show a restricted distribution in  $z$, as the young alpha sequence is concentrated to the plane. Nevertheless, even conditioned on abundances, there are the same apparent age trends seen in the left-hand panel. Old stars are preferentially seen at larger heights from the plane, and for the low-alpha sequence, very few old stars are present at the largest radial extents of the sample. For the mono-abundance selection in the panel at the far right, there are a handful of old stars present across the radial extent of the sample, preferentially at largest heights from the plane. At the same time, as also seen for the centre panel, there are young stars seen at large $z$, across all $R$. Clearly, the young-alpha sequence does not represent a homogeneously young population, and the age label demonstrates that conditioned on abundances, older stars are distributed differently to younger stars in the disk of the Milky Way.

\subsection{Stellar ages for the red giant sample}

In addition to the red clump stars, we have determined stellar masses and inferred ages (assuming the red giant evolutionary state) for the 60,000 red giant stars in DR12 which span the label-range of our training set in \teff, \logg, \feh\ and \alphafe. We include the labels for these red giant stars in Table \ref{tab:online}. 

\begin{figure}[hp]
\centering
              \includegraphics[scale=0.43]{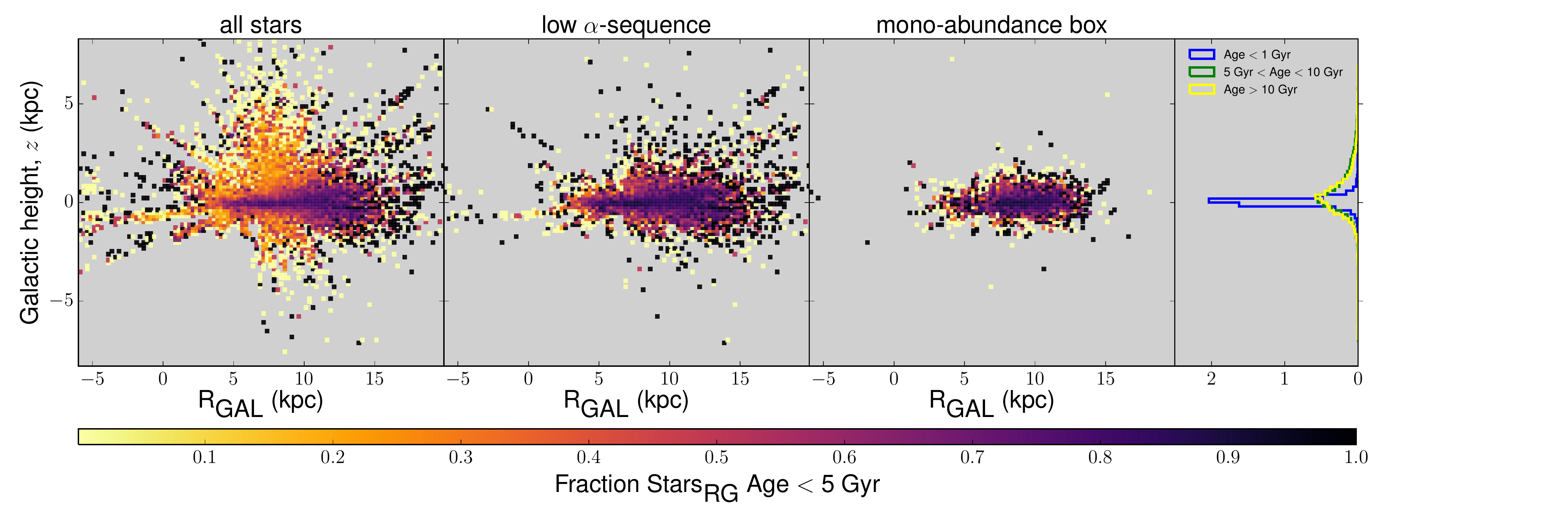}
    \caption{$(R_{GAL},z)$ Maps of the median age of the \apogee\  red giant stars showing all $\approx$ 60,000 stars at far left, the low-alpha-sequence only of these stars in the second panel from left and a small abundance bin in the low-alpha sequence third panel from left. The final panel, at right is a histogram of the different age distributions as a function of age, showing all stars across $z$, for a young, intermediate and old selection.  }
\label{fig:allage}
\end{figure}

In Figure \ref{fig:allage} we show the fractional age distribution for the 60,000 red giant stars in DR12. Note the 5\% stars where the $\chi_{reduced}^2$ $>$ 3 for \tc\ model have been removed. The distances to all of the red giant stars have been determined via interpolation to PARSEC isochrones, from the stellar parameters and by adopting the RJCE-WISE extinction value for that line of sight, provided in the \apogee\ DR12 data \citep{Majewski2011,Zasowski2013}. The panels are the same as for Figure \ref{fig:redclump_age} except now shown in terms of fractional ages (Stars with ages $<$ 5 Gyr) and for a larger extent in $(R_{GAL},z)$, as the red giant stars cover a much larger spatial region than the red clump alone. The distance uncertainty for the red giant stars is much larger than for the red clump, at about 30\%. Distances toward the bulge are particularly uncertain and likely underestimated due to the high and differential reddening in this direction (see Ness et al., 2015b, submitted).  

Figure \ref{fig:allage} demonstrates that the highest fraction of young red giant stars is in the plane of the disk and this youngest fraction flares in height with increasing radius (see the far left panel).  The stars in the outer-most region of the disk are predominantly young and stars at the solar radius at large heights from the plane are almost all old. At a given height from the plane, the stars are on average younger moving out in radius from the centre of the Galaxy. For low-alpha sequence only (middle panel), the stars toward the centre of the Galaxy comprise almost exclusively old stars and stars in the outer regions are predominantly young.  Young stars appear at all heights from the plane, even conditioned on abundances for the low-alpha sequence and the mono-abundance population. Overall the trends of the red giant sample are the same as that of the red clump sample.

\section{Discussion}

We have provided three demonstrations of the validity of the stellar masses and ages  determined with \tc. First, mathematically, \tc\ works and can return labels of \teff, \logg, \feh, \alphafe\ and mass for \apogee\ spectra, which we validate with a take-stars out test (see Figure 1). As shown with this cross-validation, we can determine log masses to an accuracy of 0.07 dex and infer log ages from these masses, to an accuracy of 0.21 dex.  

The well matched generative model to the data (see Figure \ref{fig:spectra}), from \tc's best fit labels, verifies that these five labels and polynomial model (see Section 2.1) are sufficient to very well model the behavior of the flux with the labels at test time (see Figure \ref{fig:spectra}).  In fact, the data-driven model of \tc\ trained on these five labels only (no individual abundances) provides a better match to the real data than the synthetic stellar models utilized by \aspcap.

Second, we have shown that the spectral mass (or age) indicators discovered by \tc\ are associated in the space of the actual spectra with elements that can be ``dredged-up'' (see Figure \ref{fig:mass}). Specifically, the mass information comes from the CN and CO molecules in the spectra.  Although the mass information in the \apogee\ spectral region originates from these features, in other wavelength regions it could derive from different elements or molecules. If mass information is present, it can be determined using \tc\ for other surveys, such as GALAH \citep{Freeman2012}.

Third, we show using the red clump sample of \citet{Bovy2014} that the ages of stellar structures in the Milky Way follow gross expectations, even conditioned on abundances. To demonstrate that we have a real age indicator and not a simple proxy for chemical enhancement that is tightly correlated with age (such as \alphafe) we have examined the age information within small mono-abundance boxes in \feh-\alphafe\ space for the training sample. Figure \ref{fig:alphabins} shows that there is age information within the mono-abundance bins. Furthermore, Figure \ref{fig:redclump} demonstrates the different \feh-radial profiles for the young and old red clump populations conditioned on abundances. For the low-alpha sequence only, the stars show an \feh\ distribution with radius that is consistent with radial mixing processes that are expected to be relevant for the intermediate and old populations but not the youngest stars.

The mean age map of the Milky Way disk as traced by the red clump stars shown in Figure \ref{fig:redclump_age} confirms the common wisdom that disk-thickness depends on age. Moving out in radius, younger stars are present at larger and larger heights from the plane and at small radii the youngest stars are located in significant fraction only in the plane of the disk. In the median age maps shown in Figure \ref{fig:redclump_age}, there are old stars present even for the low-alpha sequence. Therefore, the low-alpha sequence is not a homogeneously young population. The oldest stars are located preferentially at larger heights from the plane compared to the younger stars, which truncate in their distribution nearer to the plane. The age distribution trends seen in the red clump sample as a function of $(R,z)$ shown in Figure \ref{fig:redclump_age} are also seen in the red giant sample shown in Figure \ref{fig:allage}, which spans a larger extent in $(R,z)$. 

For our analysis of stellar ages presented in Figures 7 - 10, we transform our mass labels into stellar age as described in Section 4.1. Mapping the output mass labels from \tc\ to a stellar age using stellar models enforces fixed upper and lower age limits. It is also possible, however, to use \tc\ to train directly on log age rather than mass. In this case, there is no physical constraint on minimum or maximum ages, at the test step.  

We provide in Table \ref{tab:online}, in the Appendix,  a partial extract showing our stellar labels for 80,000 red giant stars from DR12 (including $\approx$ 20,000 red giant stars). This table is available in full online. We tabulate the \teff, \logg, \feh\ and \alphafe\ as well as the stellar mass label determined by \tc\ for training on log mass and  the stellar age label determined by \tc\ for the case of training directly on log age. For training on log age directly rather than log mass, the same set of reference stars are used. The label space of these stars is shown in the Appendix. The ages for the reference set of stars for training have been determined by Martig et al., (2015), who use interpolation between PARSEC isochrones, with optimized scaling relations, as a function of evolutionary state. 

 There are several promising avenues for improving our results.  Improved absolute calibrations for asteroseismic mass and radius would be highly desirable.  Our methodology would also benefit from quantifying mass loss and its stochastic uncertainty, especially for red clump stars.  A more complete stellar population study should also include corrections for the products of interacting binary star evolution, and include the impact of the IMF and star formation history on the derived mass and age distributions.  There is also the possibility of using the mass trends identified in this paper to quantify first dredge up and in situ red giant branch mixing as a function of mass and the initial abundance mixture, and to test physical theories of stellar structure and evolution.

\section*{Appendix}

\begin{figure}[h!]
\centering
        \includegraphics[scale=0.4]{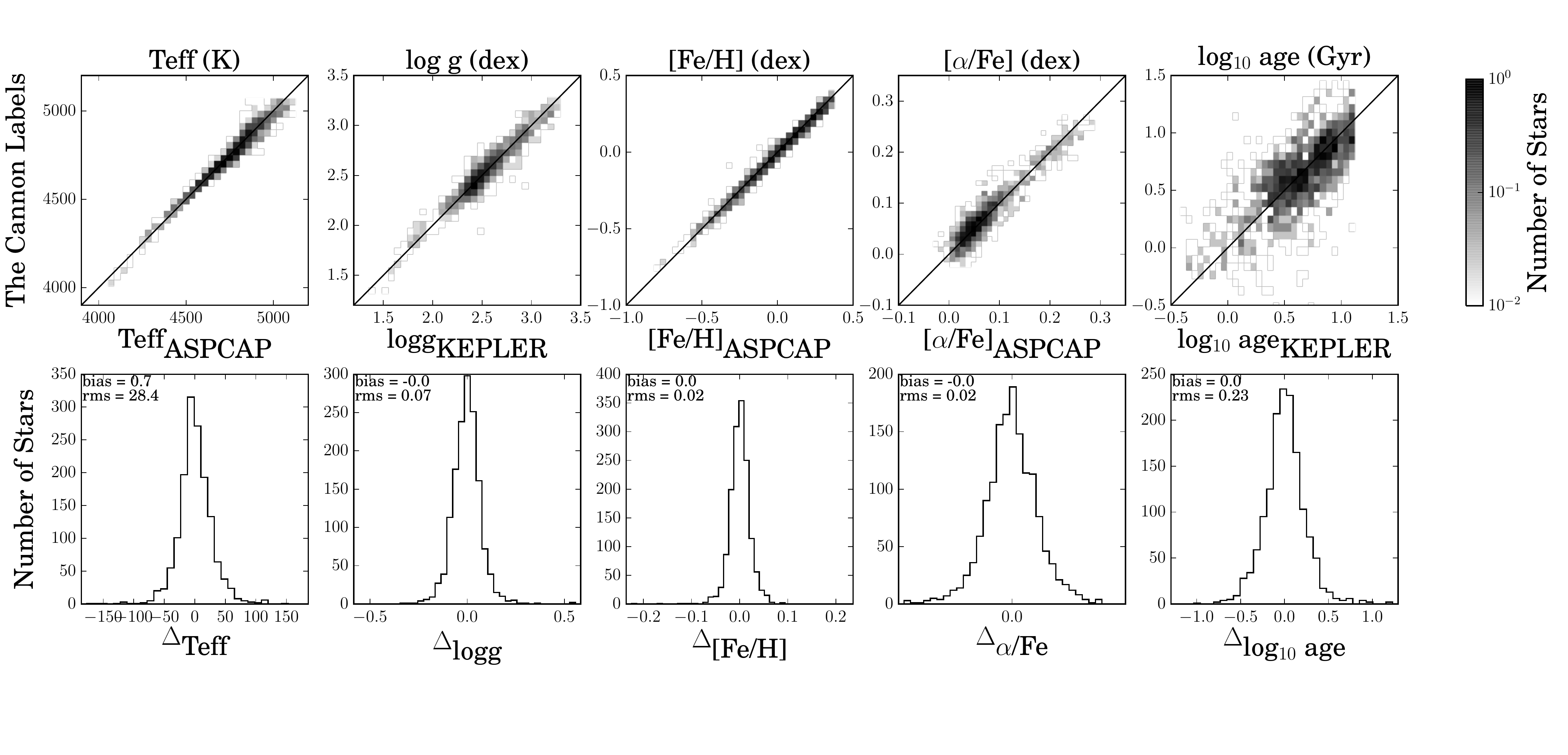}
  \caption{Cross validation of the training dataset of 1639 stars, for the \teff, \logg, \feh, \alphafe\ and age labels: the results for \tc's labels for training performed on 90\% of the \apokasc\ stars, showing the performance at test time on the 10\% of the stars not included in training, run 10 times.}
\label{fig:validation2}
\end{figure}

In Table 1, we provide the stellar parameters of \teff, \logg, \feh, \alphafe, mass and age for the DR12 red clump stars and red giant stars that are within the label range of our training set. The mass label from \tc\ is obtained for training on log mass and the age label from \tc\ is obtained for directly training on log age.  The mass label we provide can be used to infer stellar ages, using interpolation between any selected stellar isochrones and given a set of assumptions. The age inferred from the mass label from \tc, as described in Section 4.1, was used to generate the stellar ages presented in Figures 7 - 10. Training on mass and inferring age, there are stars that are artificially truncated to the maximum age from the isochrones (where masses determined by \tc\ are lower than the smallest value from the stellar evolution tracks). 

\begin{figure}[h!]
\centering
        \includegraphics[scale=0.4]{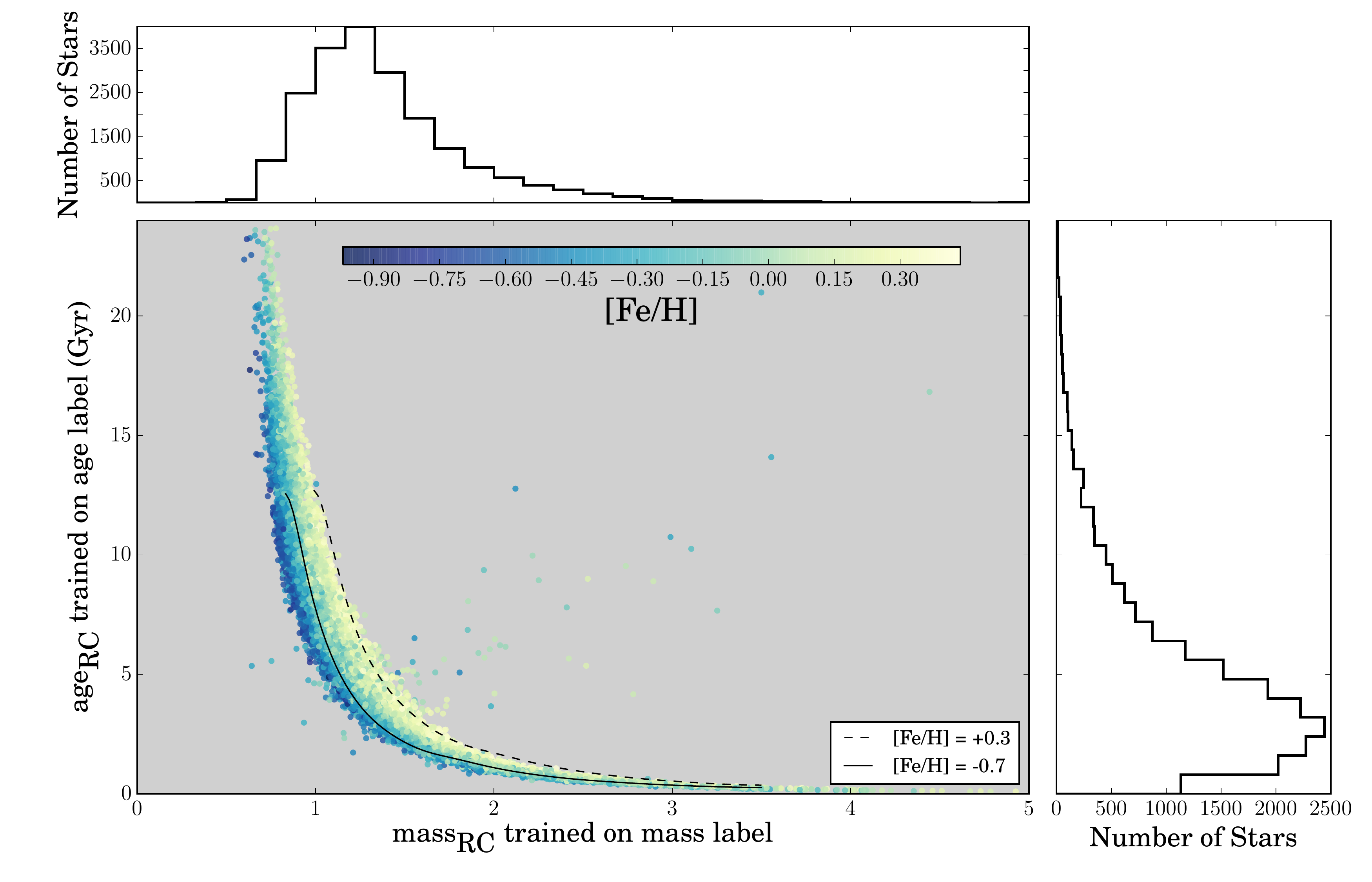}
  \caption{The mass label for the red clump sample from training on mass, compared to the age label for the red clump sample from training on age. The age distribution peaks at 2.5 Gyr for the red clump sample, which is colored by \feh. Almost all (99 percent) of the stars are within 0 $<$ age $<$ 14 Gyr. Two theoretical mass and red clump age tracks are shown, at \feh\ = +0.3 and --0.7, from PARSEC isochrones. }
\label{fig:agemass}
\end{figure}

Training on log age directly instead of log mass,  \tc\ works mathematically in the same way, as described in Section 2.1. The cross-validation result for training on age directly, instead of mass, is shown in Figure \ref{fig:validation2}. The uncertainties on the labels are similar to that of Figure 1, for training on mass. There is no physical limit in the test step of \tc\ that prohibits ages (or masses) that exceed or are smaller than that of the training set. 
\tc\ therefore is not constrained to a physically allowed regime. Therefore, training on log age results in a small subset of stars that are older than the age of the universe at the the test step, although the vast majority of stars are in physically realistic label space, 0 $<$ age $<$ 14 Gyr: only 1 percent of stars are outside of this range and typically have large associated $\chi^2_{reduced}$ values.  The comparison for the age-label determined for the red clump sample of stars, training on age, and the mass label determined for the red clump stars, training on mass, is shown in Figure \ref{fig:agemass}. Two PARSEC tracks for red clump masses and ages are shown, demonstrating that the data follow theoretical expectations.

\begin{figure}[h!]
\centering
        \includegraphics[scale=0.45]{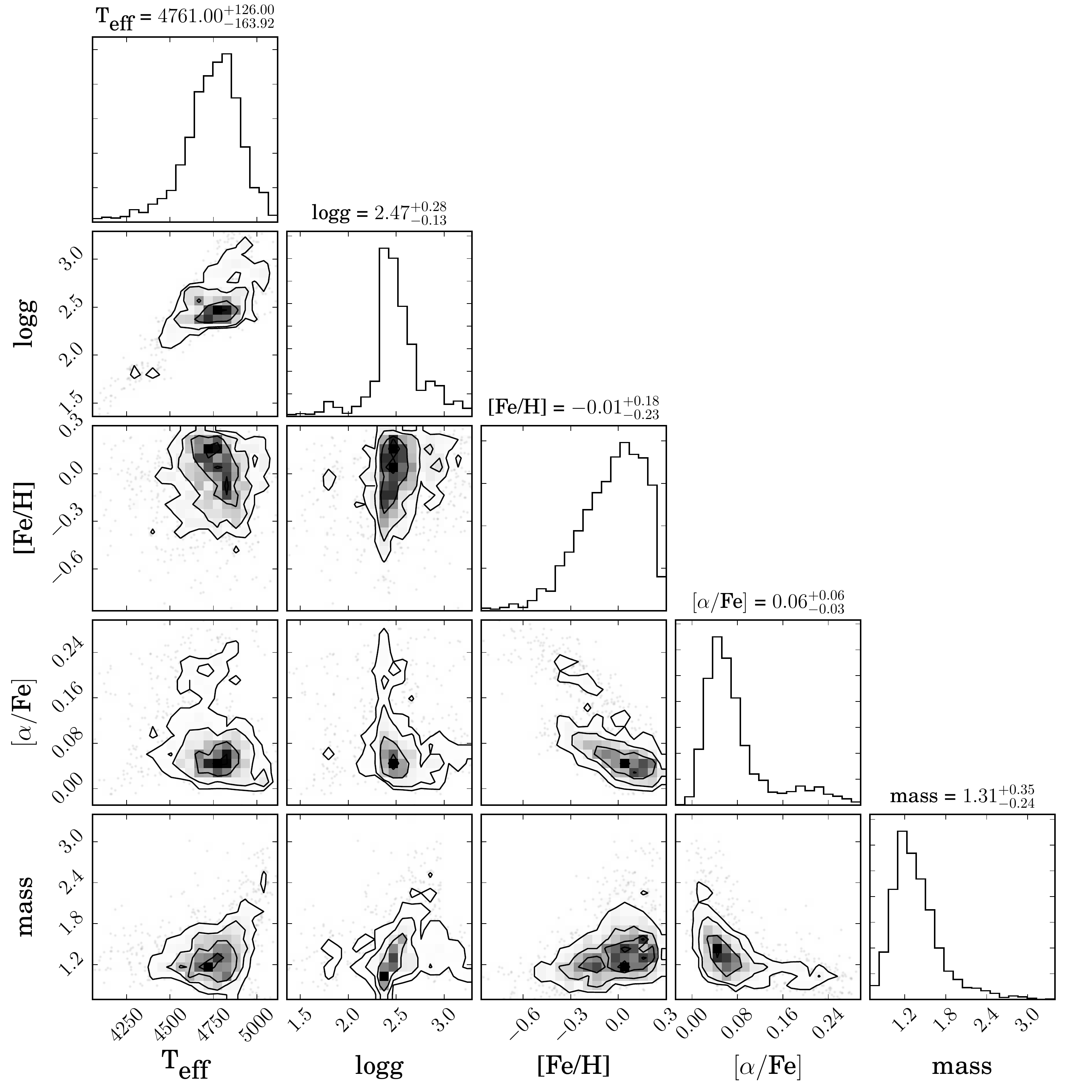}
  \caption{The label space of the training data. This Figure was made using the corner.py routine in \citet{dfm}}.
\label{fig:triangle}
\end{figure}

The label range of our training set of 1639 \apokasc\ stars is provided in Figure \ref{fig:triangle}.

Our code and documentation is located on Github. \footnote{https://github.com/mkness/TheCannon/}

\begin{table*}[h!]
\tiny{
\centering
\caption{Partial column excerpt from the online table of 6 stellar labels (\teff, \logg, \feh, \alphafe\ mass and age) determined by \tc\ for 60,000 red giant stars and 20,000 red clump stars in \apogee 's data release DR12. The errors quoted are the formal errors from \tc\ for the uncertainties on the labels (see Figures 1 and 11). The mass column in this table is for training on mass derived from seismic scaling relations and the age column in this table is derived from training on age from Martig et al., (2015)., for the same 1639 set of reference stars from \apokasc. } 

\vspace{10pt}
\begin{tabular}{| c | c | c |  c | c | c |  c | c | c | c | c | c | c | c |} 
\hline
\tiny{star ID}  & \teff\ & \logg\ & \feh\ & \alphafe\ & mass & \tiny{age} & $\sigma$(\teff) & $\sigma$(\logg) & $\sigma$(\feh) & $\sigma$(\alphafe) & $\sigma$(mass) & $\sigma$(age) & $\chi_{reduced}^2$  \\
\tiny{(2MASS)} & K & dex &  dex  & dex & $ M_{\mbox{star}}$ & Gyr & K & dex  & dex & dex &$ M_{\mbox{star}}$ & Gyr  & \\    
\hline
\tiny{21353892+4229507} & 4085.3 & 1.39  & -0.002  & 0.018 & 1.58  & 2.4 & 1.3 & 0.006 & 0.002 & 0.002 &  0.05 & 0.05 &  1.3 \\
\tiny{21354775+4233120} & 4685.8 & 1.29 & 0.070  & 0.165 & 1.02  & 10.8 &  8.3  & 0.018 & 0.006 & 0.005 &  0.02 &  0.22 & 2.3\\
\tiny{21360285+4231145} & 4493.8 & 1.72  & -0.431  & 0.025 &  1.26  & 3.5 & 5.0 & 0.018 & 0.006 & 0.005 & 0.17 & 0.6 & 2.4 \\
\tiny{21360302+4250260} & 4687.5 & 2.55  & 0.041 & 0.042 & 2.18  & 8.8 & 4.6    & 0.014 & 0.004 & 0.004 & 0.07 & 0.6 & 1.2\\
\hline
\end{tabular}
\label{tab:online} }
\end{table*}

\section*{Acknowledgments}
It is a pleasure to thank Maria Bergemann (MPIA)
  John Bochanski (Rider), Morgan Fouesneau (MPIA), Ricardo Schiavon (Liverpool John Moores University), 
  Dan Foreman-Mackey (UW), Amelia Stutz (MPIA) and Ben Weiner (Arizona State).
for valuable discussions and contributions.
This project made use of
  The NASA Astrophysics Data System,
  and open-source code in the \project{numpy} and \project{scipy} packages.
  
The research has received funding from the European Research Council under the European Union's Seventh Framework Programme (FP 7) ERC Grant Agreement n.
[321035].
  
We thank the Kavali Institute Theoretical Physics Galactic Archeology Program: this research was supported in part by the National Science Foundation under Grant No. NSF PHY11-25915.\\
 
Funding for SDSS-III has been provided by the Alfred P. Sloan Foundation, the Participating Institutions, 
the National Science Foundation, and the U.S. Department of Energy Office of Science. The SDSS-III web site is \url{http://www.sdss3.org/}.

SDSS-III is managed by the Astrophysical Research Consortium for the Participating Institutions of the SDSS-III Collaboration
 including the University of Arizona, the Brazilian Participation Group, Brookhaven National Laboratory, Carnegie Mellon University, 
 University of Florida, the French Participation Group, the German Participation Group, Harvard University, the Instituto de Astrofisica 
 de Canarias, the Michigan State/Notre Dame/JINA Participation Group, Johns Hopkins University, Lawrence Berkeley National Laboratory, 
 Max Planck Institute for Astrophysics, Max Planck Institute for Extraterrestrial Physics, New Mexico State University, New York University, 
 Ohio State University, Pennsylvania State University, University of Portsmouth, Princeton University, the Spanish Participation Group, 
 University of Tokyo, University of Utah, Vanderbilt University, University of Virginia, University of Washington, and Yale University

\end{document}